\documentclass[
a4paper,twocolumn,11pt,accepted=2022-04-08
]{quantumarticle}
\pdfoutput=1

\usepackage{hyperref}
\usepackage{amssymb,amsmath,amsfonts}
\usepackage{graphicx}
\usepackage{dcolumn,amsthm}
\usepackage{bm}
\usepackage{txfonts}
\usepackage{dsfont}
\usepackage[english]{babel}
\usepackage[utf8]{inputenc}

\newcommand{\abs}[1]{\left\lvert#1\right\rvert}
\newcommand{\norm}[1]{\lVert#1\rVert}

\providecommand{\tr}{{\rm tr}}
\renewcommand{\phi}{\varphi}
\newcommand{\bra}[1]{\left\langle #1\right\rvert}
\newcommand{\ket}[1]{\left\lvert #1\right\rangle}

\begin{document}

\title{Maximal entanglement increase  with single-photon subtraction}

\author{Kun Zhang}
\affiliation{State Key Laboratory of Precision Spectroscopy, Joint Institute of Advanced Science and Technology, School of Physics and Electronic Science, East China Normal University, Shanghai 200062, China}
\author{Jietai Jing}
\email{jtjing@phy.ecnu.edu.cn}
\affiliation{State Key Laboratory of Precision Spectroscopy, Joint Institute of Advanced Science and Technology, School of Physics and Electronic Science, East China Normal University, Shanghai 200062, China}
\affiliation{CAS Center for Excellence in Ultra-intense Laser Science, Shanghai 201800, China}
\affiliation{Department of Physics, Zhejiang University, Hangzhou 310027, China}
\affiliation{Collaborative Innovation Center of Extreme Optics, Shanxi University, Taiyuan, Shanxi 030006, China}
\author{Nicolas Treps}
\affiliation{Laboratoire Kastler Brossel, Sorbonne Universit\'{e}, CNRS, ENS-Universit\'e PSL, Coll\`{e}ge de France, 4 place Jussieu, F-75252 Paris, France}
\author{Mattia Walschaers}
\email{mattia.walschaers@lkb.upmc.fr}
\affiliation{Laboratoire Kastler Brossel, Sorbonne Universit\'{e}, CNRS, ENS-Universit\'e PSL, Coll\`{e}ge de France, 4 place Jussieu, F-75252 Paris, France}

\date{28 April 2022}

\begin{abstract}
 Entanglement is an indispensable quantum resource for quantum information technology. In continuous-variable quantum optics,  photon subtraction can increase the entanglement between  Gaussian states of light, but for mixed states the extent of this entanglement increase is poorly understood.  In this work, we use an entanglement measure based the R\'enyi-2 entropy to prove that  single-photon subtraction increases bipartite entanglement by no more than $\log 2$. This value coincides with the maximal amount of bipartite entanglement that can be achieved with one photon. The upper bound is valid for all Gaussian input states, regardless of the number of modes and the purity.
\end{abstract}

\maketitle

\section{Introduction}
Quantum entanglement \cite{AEinstein} is  commonly seen as an important resource for quantum metrology \cite{V. Giovannetti},  quantum communication \cite{C. H. Bennett2} and quantum computation \cite{P. W. Shor}.
The increase of entanglement is an important task, since it can be used to strengthen the precision of quantum metrology \cite{S. Steinlechner}, the fidelity of quantum information processing \cite{B. Schumacher,S. Lloyd}  and the efficiency of quantum computing \cite{R. Raussendorf, E. Knill}. Here, we will focus on light as a particularly interesting platform for quantum information tasks due to its intrinsic resilience against environment-induced decoherence.

Continuous-variable (CV) quantum optics \cite{S. L. Braunstein} allows for the deterministic generation of Gaussian entanglement between modes of light. In particular, it was used to produce large-scale entangled states \cite{M. Chen,S. Armstrong,X. Su,J. Roslund,S. Gerke,J.-i. Yoshikawa,K. Zhang,W. Wang}.  However, because the entangled modes are Gaussian, they do not offer a genuine quantum advantage. To implement protocols that cannot be efficiently simulated by classical computers non-Gaussian states are required \cite{S. D. Bartlett,A. Mari, S. Rahimi-Keshari}. 

A common method to produce non-Gaussian modes of light is  adding \cite{A. Zavatta} or subtracting \cite{J. Wenger, A. Ourjoumtsev,V. Parigi,Y.-S. Ra} a photon. Here, we focus on this operation's capability of increasing the entanglement between optical modes. Historically, this increase of entanglement was first seen through the study of quantum teleportation \cite{Cochrane,Olivares}. Later, more detailed studies showed that adding/subtracting a photon  to/from a two-mode squeezed vacuum state increases the states entanglement \cite{Yang Yang}. The entanglement of two-mode squeezed vacuum state was then shown to be progressively enhanced with the number of photons added (subtracted) when acting on one mode only \cite{Carlos Navarrete-Benlloch}. Similar results are obtained for single-photon addition and subtraction for pure multimode Gaussian systems \cite{M. Walschaers2,M. Walschaers1,M. Walschaers3}. This provides a way to overcome the no-go theorem which forbids entanglement distillation of Gaussian states with Gaussian measurement \cite{NoGo1,NoGo2,NoGo3}. Experimentally, this led to a demonstration of entanglement distillation on Gaussian input states \cite{EntDist}. Nevertheless, we may argue that non-Gaussian operations do not merely increase the entanglement, but also change its nature. After photon addition or subtraction, the entanglement will partially be contained in the non-Gaussian part of the state.

Non-Gaussian entanglement is notoriously challenging to study for mixed states. To deal with such states in finite-dimensional systems, negativity is perhaps the most practical measure of entanglement \cite{Vidal-Werner}. However, this measure cannot be generalised to non-Gaussian states in the infinite-dimensional Hilbert space of continuous variable systems without truncating the state to a finite number of Fock states. Entanglement of formation is another common measure of entanglement, but is also faces problems as it requires a convex roof construction and the evaluation of the von Neumann entropy \cite{R. Horodecki,C. H. Bennett,G. Toth}. In particular the latter is highly problematic for non-Gaussian states as there is no clear way to evaluate the von Neumann entropy based on phase space representations. Here, we focus on the entanglement measure based on R\'enyi-2 entropy as a viable alternative, following \cite{Kim_2010}. Because the R\'enyi-2 entropy is related to the marginal purity of the state, it can be obtained directly from the Wigner function. 

R\'enyi-2 entropy is a less popular tool than the von Neumann entropy, because, notably, it cannot be used to define mutual information for general CV states, even though Gaussian states have shown to be an exception \cite{G. Adesso1}. For Gaussian states, the entanglement measure based on R\'enyi-2 entropy give rise to monogamy relations \cite{Lami1} and operationally it appears as a natural upper bound on the secret key distillation rate when Alice and Bob are restricted to Gaussian measurements \cite{Lami2}. Many of these attractive features of R\'enyi-2 entropy cannot be generalised to non-Gaussian states. Yet, the entanglement measure of \cite{Kim_2010} remains well-defined for all CV states, and we will focus on it throughout the remainder of this manuscript.

\section{Continuous-variable systems}

We study CV systems, which can be though of as ensembles of quantum harmonic oscillators \cite{ModesStates}. When the system has $m$ modes, we can define a mode basis with $m$ associated creation and annihilation operators labeled $\hat a^{\dag}_1, \dots, \hat a^{\dag}_m$ and $\hat a_1, \dots, \hat a_m$, respectively. These operators satisfy the canonical commutation relations
\begin{equation}
    [\hat a_j, \hat a^{\dag}_k] = \delta_{j,k}, \quad [\hat a_j, \hat a_k]  = [\hat a^{\dag}_j, \hat a^{\dag}_k] = 0.
\end{equation}
Furthermore, superpositions of modes define new modes, which implies that for any set of coefficients $c_j \in \mathbb{C}$ with $\sum_{j=1}^m \abs{c_j}^2 = 1$,
\begin{equation}\label{eq:submode}
    \hat a' = \sum_{j=1}^m c_j \hat a_j
\end{equation}
is a well-defined annihilation operator and $[\hat a', \hat { a'}^{\dag}] = 1$.

When studying the CV aspects of such systems, it is common to use quadrature operators instead. In our convention, these are defined as
\begin{align}
    &\hat x_j = \hat a^{\dag}_j + \hat a_j ,\\
    &\hat p_j = i(\hat a^{\dag}_j - \hat a_j).
\end{align}
The canonical commutation relations then translates to $[\hat x_j, \hat p_k] = 2i\delta_{j,k}$. The advantage of using quadrature operators is that they are observables that can can be measured. In multimode quantum optics, for example one uses homodyne detection to measure these observables. The possible measurement outcomes of these quadratures are real and form a continuum --hence CV-- and can be grouped together in a $2m$ dimensional optical phase space.

this phase space provides possibilities to represent any quantum state $\hat \rho$ in terms of a quasi-probability distribution \cite{Tutorial}. In this work, we rely the Wigner function, which is given by
\begin{equation}\label{eq:DefWig}
    W(\vec \beta) = \frac{1}{(2\pi)^{2m}}\int_{\mathbb{R}^{2m}} \tr[\hat \rho e^{i\vec\gamma^{\top}\vec{\hat r}}] e^{-i\vec\gamma^{\top}\vec \beta},
\end{equation}
where $\vec{\hat r}$ is a vector of operators, given by $\vec{\hat r} = (\hat x_1, \dots, \hat x_m, \hat p_1, \dots, \hat p_m)^{\top}$, and $\vec \beta$ is a vector in phase space that can be explicitly represented as $\vec \beta = (x_1, \dots, x_m, p_1, \dots, p_m)^{\top}.$
The Wigner function is particularly useful since its marginals correspond to measurement statistics of the quadrature observables and furthermore it also allows for a direct calculation of purity, which we exploit in the following sections.

\section{R\'enyi-2 entanglement}

Throughout the work, we will restrict ourselves to bipartite entanglement. As shown in Fig.~\ref{fig1}(a), for  an arbitrary  global pure state $\hat \rho=|\psi\rangle\langle\psi|$, the entanglement  between two subsystems $\mathcal{A}$ and $\mathcal{B}$  
can be measured through the R\'enyi-$2$ entropy as
\begin{equation}\label{entMeasureDef}\mathcal{E}_{R}(|\psi\rangle) =-\mathrm{log}\mu(\hat{\rho}^{}_{\mathcal{A}}),
\end{equation}
where $\hat{\rho}_{\mathcal{A}} = \mathrm{tr}_{\mathcal{B}} \hat \rho$ is the reduced density operator for subsystem $\mathcal{A}$. The measure is based on the marginal purity of one of the two subsystems $\mu_{\mathcal{A}}=\mathrm{tr}(\hat{\rho}_{\mathcal{A}}^{2})$,  which can directly be calculated from the Wigner function \eqref{eq:DefWig} by \begin{equation}\label{eq:purityFromWigner}\mu_{\mathcal{A}}=(4\pi)^{{m_{\mathcal{A}}}}\int_{\mathbb{R}^{2{m_{\mathcal{A}}}}}|W_{\mathcal{A}}(\vec\beta)|^2 \mathrm{d}^{2{m_{\mathcal{A}}}}\vec\beta,\end{equation} where ${m_{\cal A}}$ is the total number of modes in subsystem $\mathcal{A}$. Importantly, this method works both for Gaussian and non-Gaussian states. Its extension to mixed states can be obtained by the following convex construction:
\begin{equation}\label{eq:ConvexRoof}
    {\cal E}_R(\hat\rho) = \inf_{\{p(\lambda), \ket{\psi_{\lambda}}\}} \int {\rm d}\lambda\, p(\lambda) {\cal E}_R(\ket{\psi_{\lambda}}),
\end{equation}
where we minimize over all decompositions $\hat\rho = \int {\rm d}\lambda\, p(\lambda) \ket{\psi_{\lambda}}\bra{\psi_{\lambda}}$. Because the the R\'enyi-2 entropy is a lower bound on the von Neumann entropy, it naturally follows that ${\cal E}_R(\hat\rho)$ is a lower bound on the entanglement of formation. Nevertheless, it is also a meaningful entanglement measure in its own right, which we will use to derive an intuitive upper limit of the entanglement increase: we will show that for any Gaussian state --mixed or pure-- the maximal increase of bipartite ``R\'enyi-2 entanglement'' through single-photon addition/subtraction is $\log 2$. \\



\section{Photon subtraction}
\begin{figure}
\centering
\includegraphics[scale=0.35]{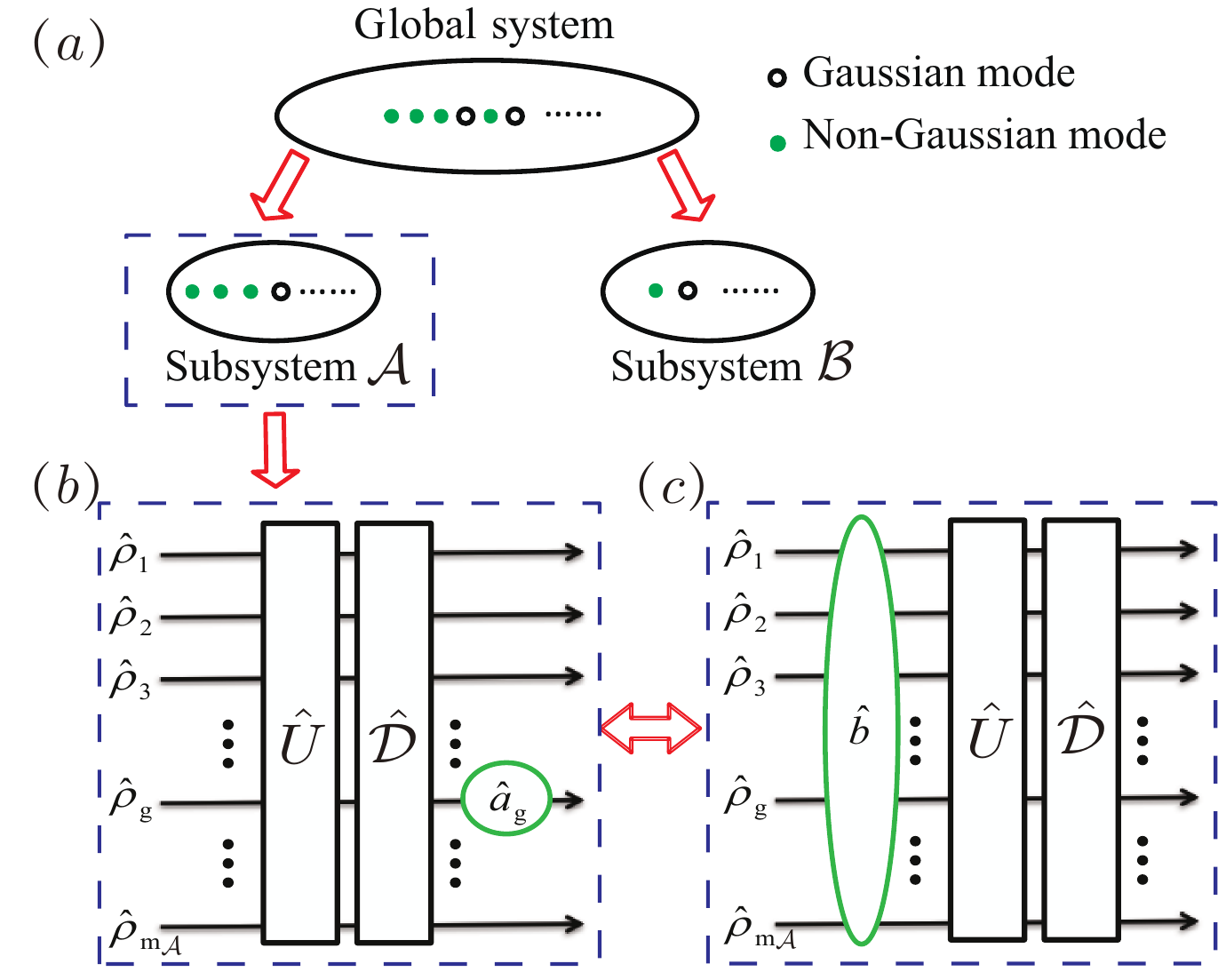}\\
\caption{\footnotesize (a) Schematic diagram of assigning all modes in a global pure state to subsystems ${\cal A}$ and ${\cal B}$. The green dots refer to non-Gaussian modes, and the small black circles refer to Gaussian modes.   (b) Thermal decomposition of subsystems $\mathcal{A}$, in which  a
photon is subtracted (surrounded by a green circle) in the mode-$g$ after the displacement operator $\mathcal{\hat{D}}$.  (c)  A combined operator $\hat{b}$ (surrounded by a green ellipses), which  adds and subtracts a photon  to and from each thermal mode  before the unitary operator $\hat{U}$. The operations in (b) and (c) are different, but the Bogoliubov transform tells us that the ultimate output state is strictly the same.
} \label{fig1}
\end{figure}

We now focus on entanglement increase via photon subtraction. This operation is theoretically described by the action of an annihilation operator on the (in our case Gaussian) state. We assume that this operation is implemented on subsystem $\mathcal{A}$, as shown in Fig.~\ref{fig1}. To study the increase of entanglement, we start by analysing the marginal purity of the photon subtracted state in subsystem $\mathcal{A}$:
\begin{eqnarray}\label{eq:sub}
\hat{\rho}^{-}_{\mathcal{A}}=\frac{\hat{a}_{g}\hat{\rho}_{\mathcal{A}}\hat{a}_{g}^{\dag}}{\mathrm{tr}(\hat{a}_{g}^{\dag}\hat{a}_{g}\hat{\rho}_{\mathcal{A}})}.
\end{eqnarray}
The label $g$ indicates the mode in which the photon is subtracted. Using \eqref{eq:submode} we can take $\hat{a}_{g}$ to be an annihilation operator acting on an arbitrary superposition of modes. Our main tool to evaluate the marginal purity of the state (\ref{eq:sub}) are Gaussian transformations \cite{C. Weedbrook}. As shown in Fig. \ref{fig1}(b), we can write a thermal decomposition form for an arbitrary subsystem $\mathcal{A}$ as follows
\begin{eqnarray}\label{00main}
\hat{\rho}_{{\mathcal{A}}}=\mathcal{\hat{D}}\hat{U}\mathop{\otimes}\limits_{i=1}^{{m_{A}}}\hat{\rho}_{i}\hat{U}^{\dag}\mathcal{\hat{D}}^{\dag},
\end{eqnarray}
where $\hat{U}$ is a Bogoliubov transformation and $\mathcal{\hat{D}}=\prod\limits_{i=1}^{m_\mathrm{A}}\hat{D}(\alpha_{i})$ is a displacement operator. The unitary displacement operators $\hat{D}(\alpha_{i})$ change the mean field of the state in mode $i$ and are given by $\hat{D}(\alpha_{i}) \coloneqq \exp\left[i (\alpha_{i}\hat a^{\dag}_i + \alpha^*_{i} \hat a_i)\right]$
such that its action on the annihilation operators is given by $\hat{D}^{\dag}(\alpha_{i})\hat a_i\hat{D}(\alpha_{i})= \hat a_i +\alpha_i$. It is now convenient to define $\vec{\hat{a}}^{\dag}=(\hat{a}_{1}^{\dag},\cdots,\hat{a}_{{m_{\mathcal{A}}}}^{\dag})^{\top}$ and  $\vec{\hat{a}}=(\hat{a}_{1},\cdots,\hat{a}_{{m_{\mathcal{A}}}})^{\top}$. As such, we find 
\begin{equation}\label{eq:D}
\mathcal{\hat{D}}^{\dag}\vec{\hat{a}}\mathcal{\hat{D}} = \vec{\hat{a}} + \vec \alpha.
\end{equation}
In a similar spirit, a Bogoliubov transformation is generated by a Hamiltonian that is quadratic in the creation and annihilation operators. However, the simplest way to define it is through its action on creation and annihilation operators
\begin{equation}\label{eq:U}
    \hat{U}^{\dag}\vec{\hat{a}}\hat{U} = K\vec {\hat{a}}^{\dag} + L \vec{\hat{a}},
\end{equation}
with $L^{\dag}L - K^{\dag}K = \mathds{1}$ and $K^{\dag}L = -L^{\dag}K$. The specific details of the Bogoliubov transformation $\hat{U}$ depend on the state $\hat{\rho}_{{\mathcal{A}}}$ and are dictated by the Williamson decomposition \cite{C. Weedbrook}. 

Furthermore, single-mode thermal states $\hat{\rho}_i$ can be fully characterised by the amount of thermal noise in units of shot noise $n_i$.  
Since the Gaussian unitary $\mathcal{\hat{D}}\hat{U}$ does not affect the marginal purity of subsystem $\hat{\rho}_{{\mathcal{A}}}$  can be directly expressed as \cite{C. Weedbrook} 
\begin{equation}
\mu_{{\mathcal{A}}}  = \frac{1}{\prod\limits_{i=1}^{m_\mathcal{A}}n_i}.
\end{equation}
If the global state is pure, this marginal purity can directly be connected to the amount of entanglement in the initial Gaussian state through \eqref{entMeasureDef}.


As shown  by the green circles in Fig. \ref{fig1}(b) and Fig. \ref{fig1}(c), if we subtract a photon from the mode $g$ of subsystem ${\mathcal{A}}$,
we can use  Bogoliubov transform to convert the photon-subtraction operation at mode $g$ into  a combined operator $\hat{b}$, which  adds and subtracts a photon on each thermal mode
\begin{equation}\begin{split}
    \label{parallelmain}
&\hat{\rho}_{{\mathcal{A}}}^{-}=\frac{\mathcal{\hat{D}}\hat{U}\hat{b}\mathop{\otimes}\limits_{i=1}^{{m_{\mathcal{A}}}}\hat{\rho}_{i}\hat{b}^{\dag}\hat{U}^{\dag}\mathcal{\hat{D}}^{\dag}}{\mathrm{tr}(\hat{b}^{\dag}\hat{b}\mathop{\otimes}\limits_{i=1}^{{m_{\cal A}}}\hat{\rho}_{i})},\\ &\text{with }\quad \hat{b} = \hat{U}^{\dag}\mathcal{\hat{D}}^{\dag}\hat{a}_{g}\mathcal{\hat{D}}\hat{U}.\end{split}
\end{equation}
We can combine \eqref{eq:D} and \eqref{eq:U} to explicitly write that $\hat{b}=\vec k \cdot \vec {\hat{a}}^{\dag}+\vec l \cdot \vec{\hat{a}} +\alpha_{g},$
where $\vec k=(k_{1},\cdots,k_{{m_{\mathcal{A}}}})$ and $\vec l=(l_{1},\cdots,l_{{m_{\mathcal{A}}}})$ are ${m_{\mathcal{A}}}$-dimensional complex vectors, and $\alpha_g$ is a complex number, all depending on the specific choice of mode $g$.

\section{Relative marginal purity}
Based on both the Williamson decomposition (\ref{00main}) and Bogoliubov transformation (\ref{parallelmain}), we can  derive (see Appendices \ref{A1} and \ref{B1}) a general expression of relative marginal purity
\begin{eqnarray}\label{RP1main}
 \nonumber \frac{\mu_{{\mathcal{A}}}^{-}}{\mu_{{\mathcal{A}}}} =&&\frac{1}{2}+\Bigg[\frac{1}{2}(\sum_{i=1}^{{m_{\mathcal{A}}}}\frac{\tilde{{N}}_{i}}{n_i})^2+
 \frac{1}{2}|\alpha_{g}|^4+|\sum_{i=1}^{{m_{\mathcal{A}}}}k_{i}l_{i}\frac{n_i^2-1}{2n_{i}}|^2\\
 \nonumber &&+{\alpha_{g}^{\ast}}^2(\sum_{i=1}^{{m_{\mathcal{A}}}}\frac{k_{i}l_{i}(n_{i}^{2}-1)}{2n_{i}})+\alpha_{g}^2(\sum_{i=1}^{{m_{\mathcal{A}}}}\frac{k_{i}^{\ast}l_{i}^{\ast}(n_{i}^{2}-1)}{2n_{i}})\\
 &&+|\alpha_{g}|^2\sum_{i=1}^{{m_{\mathcal{A}}}}N_{i}\Bigg]/{(\sum_{i=1}^{{m_{\mathcal{A}}}}{N}_{i}+|\alpha_{g}|^2)^2},
\end{eqnarray}
where ${N}_{i}=|k_{i}|^2\frac{n_i+1}{2}+|l_{i}|^2\frac{n_i-1}{2}$, and ${\tilde{N}}_{i} =|k_{i}|^2\frac{n_i+1}{2}-|l_{i}|^2\frac{n_i-1}{2}.$
By minimising this expression, we find that  \begin{equation}\label{eq:purity}\frac{\mu_{{\mathcal{A}}}^{-}}{\mu_{{\mathcal{A}}}}\geqslant{1/2}.\end{equation}
We thus show that photon subtraction decreases the marginal purity of an arbitrary state by at most a factor two.  In the examples at the end of this Article, we show that this lower bound is tight in the sense that there are states for which ${\mu_{{\mathcal{A}}}^{-}}={\mu_{{\mathcal{A}}}}/2$. Furthermore, when there is no entanglement in the initial Gaussian state, i.e. when $\mu_{\cal A} = 1$, we find that $\mu^{-}_{\cal A} = 1$.\\

\section{Bounds on the increase of R\'enyi-2 entanglement}

When the state for the global system is pure, we can immediately combine the definition (\ref{entMeasureDef}) with (\ref{eq:purity}) to obtain that the entanglement increase is governed by
\begin{equation}\label{eq:entred}
    \Delta {\cal E}_R = \log \mu_{\cal A} - \log \mu^-_{\cal A} \leqslant \log 2.
\end{equation}
For mixed states, on the other hand, we use the definition of (\ref{eq:ConvexRoof}), which implies that we must consider all possible decompositions of the state. For each state in the decomposition we can use (\ref{eq:purityFromWigner}) to calculate $\mathcal{E}_{R}(|\psi_{\lambda}\rangle)$ based on the Wigner function. This construction naturally implies that
$\mathcal{E}_{R}(\hat{\rho}^G) \leqslant \int {\rm d}\lambda p^{\prime}(\lambda) \mathcal{E}_{R}(|\psi_{\lambda}\rangle)$, 
for any possible decomposition of $\hat{\rho}$ in pure states.  
For Gaussian states this idea was explicitly used to define Gaussian
entanglement of formation \cite{M. M.Wolf} as an upper bound to the entanglement of formation. A similar idea was later used to define the Gaussian R\'enyi-2 entanglement \cite{G. Adesso1}. Here, we generalise this idea to the class of photon-subtracted states by using that every mixed photon-subtracted state can be decomposed as a mixture of pure photon-subtracted states.
  
 The Wigner function \cite{C. Weedbrook}  of a Gaussian state $\hat\rho^G$ can be  written   as
\begin{eqnarray}\label{Wg}
 W^{G}(\vec \beta) &=& \frac{\mathrm{exp}[{-\frac{1}{2}(\vec\beta-\vec\alpha_0)^{\top}V^{-1}(\vec\beta-\vec\alpha_0)}]}{{(2\pi)^{{m}}\sqrt{\mathrm{det}(V)}}},
\end{eqnarray}
where $\vec{\beta}=({x}_1, \cdots, {x}_{m},$ ${p}_1, \cdots, {p}_{m})^{\top}\in\mathbb{R}^{2{m}}$  is a set of
amplitude and phase quadrature and  $\xi$   is the corresponding displacement vector. 
And the covariance matrix $V$  can be decomposed as $V = V_{p} +V_{c} $,  where $V_{p}$ is the covariance matrix of a pure state, and $V_{c}$ is some positive-definite matrix that describes additional classical noise. This decomposition is generally not unique, but we can use a constructive technique to prove that this decomposition always exists \cite{M. Walschaers1}.  
We can think of the state $\hat{\rho}^G$ as
being generated by injecting a pure state $\hat{\rho}_{p}$   into a noisy Gaussian channel, so that
the Gaussian state $\hat \rho^G$ can be decomposed \cite{M. M.Wolf} as
\begin{eqnarray}\label{E6}
 \hat\rho^G &=& \int_{\mathbb{R}^{2m}} \mathrm{d}^{2m}\vec\alpha\, \mathcal{\hat{D}}(\vec\alpha)\hat{\rho}^G_{p}\mathcal{\hat{D}}^{\dag}(\vec\alpha) p_c^{G}(\vec \alpha),
\end{eqnarray}
where $\hat{\rho}_{p}$ is a pure squeezed vacuum state with covariance matrix $V_p$. We also introduce the probability distribution
\begin{equation}
    p_c^{G}(\vec \alpha)=\frac{e^{{-\frac{1}{2}(\vec\alpha-\vec\alpha_0)^{\top}V_{c}^{-1}(\vec\alpha-\vec\alpha_0)}}}{{(2\pi)^{{m}}\sqrt{\mathrm{det}(V_{c})}}},
\end{equation}with $\vec \alpha_0$ the mean field of $\hat\rho^G$. This gives us an explicit way to decompose any  Gaussian state as a mixture of squeezed vacuum states.
This decomposition of the Gaussian state can be used to decompose the associated photon-subtracted state as
\begin{eqnarray}\label{eq:decompSub}
 \hat\rho^{-} &=& \int_{\mathbb{R}^{2m}} \mathrm{d}^{2m}\vec \alpha \hat{\rho}_{p,\vec \alpha}^{-} p_{c}^{-}(\vec \alpha),
\end{eqnarray}
where the pure state $\hat{\rho}_{p,\alpha}^{-}$ is the state that is obtained by subtracting a single photon from $\mathcal{\hat{D}}(\alpha)\hat{\rho}_{p}\mathcal{\hat{D}}^{\dag}(\alpha)$, and $p^-_c(\alpha)$ a probability distribution that is explicitly given by \cite{M. Walschaers1}
\begin{equation}\label{eq:pc} \begin{split}p_{c}^{-}(\vec\alpha)=&\left(\frac{\tr [V_{p; g}] + \norm{\vec \alpha_g}^2 - 2}{\mathrm{tr} [V_g] + \norm{\vec\alpha_{0;g}}^2 - 2}\right)\\
&\times\frac{e^{[{-\frac{1}{2}(\vec\alpha-\vec\alpha_0)^{\top}V_{c}^{-1}(\vec\alpha-\vec\alpha_0)}]}}{{(2\pi)^{{m}}\sqrt{\mathrm{det}(V_{c})}}},\end{split}\end{equation}
where we introduce $V_g$ and $V_{p;g}$ as the $2 \times 2$ sub-matrices of $V$ and $V_{p}$, respectively, that describe the mode $g$. Analogously, the $2$-dimensional vectors $\vec \alpha_g$ and $\vec \alpha_{0;g}$ describe the projections of $\vec \alpha$ and $\vec \alpha_0$, respectively, to mode $g$.

The definition \eqref{eq:ConvexRoof} imposes to take the infimum over all possible decompositions of $\hat\rho^{-}$ in terms of pure states. This implies that the average entanglement of the pure state in any possible decomposition is higher than ${\cal E}_{R}(\hat\rho^{-})$. The decomposition (\ref{eq:decompSub}) can thus be used to derive an upper bound for the entanglement increase, optimised over all possible decompositions of $V$. This leads to a first important result, we find that 
\begin{equation}\label{eq:decompEnt1}{\cal E}_{R}(\hat\rho^{-}) \leqslant \inf_{V_p \leqslant V} \int_{\mathbb{R}^{2m}} \mathrm{d}^{2m}\vec\alpha p_{c}^{-}(\vec \alpha)\mathcal{E}_{R}(\hat{\rho}_{p,\vec \alpha}^{-}).
\end{equation}
We may then use (\ref{eq:pc}) to explicitly bound ${\cal E}_{R}(\hat\rho^{-})$ by using only the initial state's covariance matrix $V$, the mean field $\vec \alpha_0$, and the subtraction mode $g$. Still, approaching the infimum in (\ref{eq:decompEnt1}) remains a challenging task, which we will not explore in detail in this Article.

Rather than looking for state-dependent bounds, we now focus on a universal bound that quantifies the best possible entanglement increase through photon subtraction. Because $\hat{\rho}_{p,\alpha}^{-}$ is a single-photon subtracted pure state, we can use (\ref{eq:entred}) to obtain that ${\cal E}_{R}(\rho^{-}_{p, \vec\alpha}) \leqslant \log 2 + {\cal E}_{R}(\mathcal{\hat{D}}(\vec \alpha)\hat{\rho}_{p}\mathcal{\hat{D}}(\vec\alpha)^{\dag})$. Furthermore, because entanglement is unchanged under local unitary transformations, we find that ${\cal E}_{R}(\mathcal{\hat{D}}(\vec\alpha)\hat{\rho}_{p}\mathcal{\hat{D}}(\vec\alpha)^{\dag})= {\cal E}_{R}(\hat{\rho}_{p})$. Inserting all these elements in (\ref{eq:decompEnt1}), we find that 
\begin{equation}
   {\cal E}_{R}(\hat\rho^{-}) \leqslant  \log 2 +\inf_{V_p \leqslant V}[\mathcal{E}_{R}(\hat{\rho}_{p})].
\end{equation}
The second term of the right side is by definition the Gaussian R\'enyi-2 entanglement of the mixed state $\hat\rho^G$ \cite{G. Adesso1}. This entanglement measure is defined by minimising over all possible Gaussian decompositions, but it is a priori not clear that there cannot be any decomposition in non-Gaussian states that yields a lower entanglement. In other words, we know that $\inf\limits_{V_p \leqslant V}\mathcal{E}_{R}(\hat{\rho}_{p}) \leqslant \mathcal{E}_{R}(\hat{\rho}^G)$, but it is not clear whether or not this is actually an equality. 

Hence,  we have shown that photon subtraction on any arbitrary Gaussian state $\hat\rho^G$ can increase the R\'enyi-2 entanglement by at most $\log 2$ as compared to its Gaussian R\'enyi-2 entanglement. A completely analogous calculation for photon-added state results in the same bound even though photon subtraction and photon addition tend to increase the entanglement by different amounts depending on the state.\\

   \begin{figure}
\centering
\includegraphics[scale=0.45]{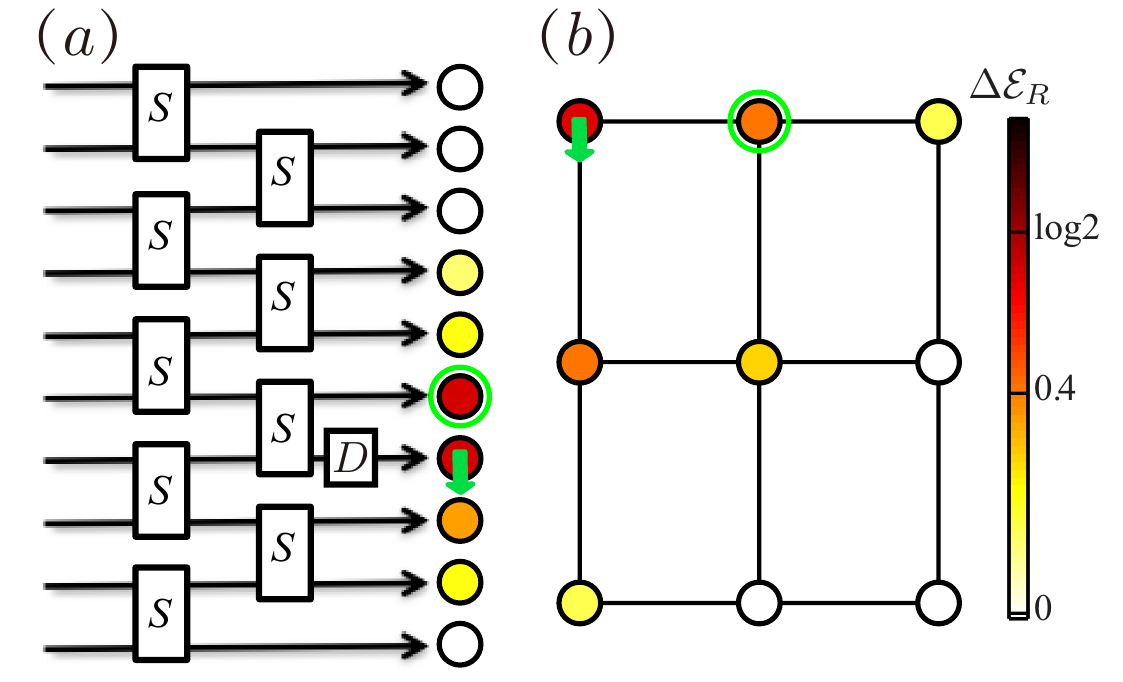}
\caption{\footnotesize Two examples of multimode entangled pure states. The optical modes are indicated by black circles,  in which the color code  indicates the increase (${\Delta\cal E}_{R}$) in  R\'enyi-2 entanglement between the corresponding mode and the rest of the system. The mode $g$ with one photon subtracted is indicated by  a downward green arrow and its neighboring mode, mode $g^{\prime}$, are marked with a green circle.  
(a) is a parallel optical network with 10 modes, and  all these modes are in the initial  vacuum sates. $S$ indicates the two-mode squeezer operator with parameter  $r=1$.  $D$ is the displacement operator, which makes mode  $g$ produce a displacement of $\alpha_g=0.5$.  (b) is a $3\times3$ graph state.  All these modes are in the initial squeezed vacuum with $10$ dB of squeezing. And the mode $g$ also has  a displacement of $\alpha_g=0.5$. 
} \label{fig2}
\end{figure}

\section{Examples}
To get a better understanding of the quality of the bound and its physical interpretation, we will consider a range of examples. Because there is no exact method to calculate the entanglement for non-Gaussian mixed states, we restrict ourselves to pure states where we can use (\ref{eq:entred}) to calculate the entanglement increase. We consider the following two typical entangled pure states as examples:
\begin{enumerate}
   \item We subtract a photon from a linear entangled network, which is composed of several  two-mode squeezer gates, as shown in Fig. \ref{fig2}(a). All the modes are initially in  vacuum states $\hat{\rho}_{\mathrm{o}}$,  then the output state can be expressed as $\hat{\rho}=\hat{\mathcal{D}}\hat{U}\hat{\rho}_{\mathrm{o}}\hat{U}^{\dag}{\hat{\mathcal{D}}^{\dag}}$, where $\hat{U}=\hat{S}_{m-1}\cdots\hat{S}_1$ with $\hat{S}_{i}=\mathrm{exp}[r(\hat{a}_{i}\hat{a}_{i+1}-\hat{a}_{i}^{\dag}\hat{a}_{i+1}^{\dag})/2]$. To test the effect of a mean field, we add the operator $\hat{\mathcal{D}}$ that adds a displacement $(\alpha_{g})$ to the mode-g. Thus the covariance matrix of the state $\hat{\rho}$ is $V_{\mathrm{network}}=C^{t}C$, where  $C$ is a symplectic matrix that implements the multimode squeezing operation on the quadratures $\hat{U}^{\dag}\hat{\beta}\hat{U} = C\hat{\beta}$. 
    \item We  subtract a photon from  a graph state, as shown in Fig. \ref{fig2}(b), which is the backbone of measurement-based CV quantum computation. We start out from a set of independent squeezed modes, with a joint covariance matrix given by $V_{0}=\mathrm{diag}(s,\cdots,s,s^{-1},\cdots,s^{-1})$. Every edge of
the graph corresponds to a $C_Z$ gate, which turn those non-correlated modes  into a graph state. On the level of covariance matrices, the action of the set of $C_Z$ gates is represented by
a symplectic transformation $C$, applied to $V_0$ \cite{M. Walschaers3}:
 $V_{\mathrm{graph}}=C^{t}V_0C$, with $ C=\left(
                                              \begin{array}{cc}
                                                \mathds{1} & \mathbb{A}\\
                                                0 & \mathds{1} \\
                                              \end{array}
                                            \right)$,
where $\mathbb{A}$ is $m\times m$ adjacency matrix.
\end{enumerate}

We can use methods purely based on phase space representations \cite{M. Walschaers1,M. Walschaers2} to describe the Wigner function of a reduced photon subtracted state for any subsystem ${\cal A}$. A very detailed derivation can be found in \cite{Tutorial}, leading to the general expression
\begin{eqnarray}\label{wngmain}
 W^{-}_{\cal A}(\vec\beta_{\cal A})=&& \Big(\norm{X V^{-1}_{\cal A}(\vec\beta_{\cal A} - \vec\alpha_{\cal A}) - \vec\alpha_g}^2\\ 
&&\quad+ \tr(V_g - X V^{-1}_{\cal A} X^{\top}) - 2 \Big)\nonumber\\
&&\times \frac{W^{G}_{\cal A}(\vec\beta_{\cal A})}{{\norm{\vec\alpha_g}^2 + \tr(V_g) - 2}},\nonumber
\end{eqnarray}
where $\vec{\beta}_{\cal A}\in\mathbb{R}^{2{m_{\cal A}}}$ and $X =  G^{\top}(V - \mathds{1})A$. Here, $G$ is a $2m \times 2$ matrix where the two columns are the basis vectors $\vec g^{(x)}$ and $\vec g^{(p)}$ associated with the two phase space axes of mode $g$:
\begin{equation}
    G = \begin{pmatrix} \mid & \mid \\
    \vec g^{(x)}& \vec g^{(p)}\\
    \mid & \mid
    \end{pmatrix}.
\end{equation}
Analogously, $A$ is a $2m \times 2m_{\cal A}$ matrix where the columns are the symplectic basis vectors $\vec a^{(x)}_1, \dots ,\vec a^{(x)}_{m_{\cal A}},\vec a^{(p)}_1, \dots ,\vec a^{(p)}_{m_{\cal A}}$ that generate the phase space of subsystem $\cal A$:
\begin{equation}
    A = \begin{pmatrix} \mid & &\mid & \mid & & \mid \\
    \vec a_1^{(x)}& \cdots & \vec a_{m_{\cal A}}^{(x)}&\vec a_1^{(p)}& \cdots & \vec a_{m_{\cal A}}^{(p)}\\
    \mid & & \mid & \mid & & \mid
    \end{pmatrix}.
\end{equation}
In (\ref{wngmain}), we then obtain the matrices $V_{\cal A} = A^{\top}VA$ and $V_g = G^{\top}VG$, and the displacement vectors $\vec \alpha_{\cal A} = A^{\top}\vec \alpha$ and $\vec \alpha_{g} = G^{\top}\vec \alpha$. We can now use standard techniques of Gaussian integrals to evaluate the marginal purity (\ref{eq:purityFromWigner}) from $W^{-}_{\cal A}(\vec\beta_{\cal A})$ for an arbitrary bipartition ${\cal A}$. This in turn allows to directly evaluate the entanglement increase (\ref{eq:entred}).





\begin{figure}
\centering
\includegraphics[scale=0.45]{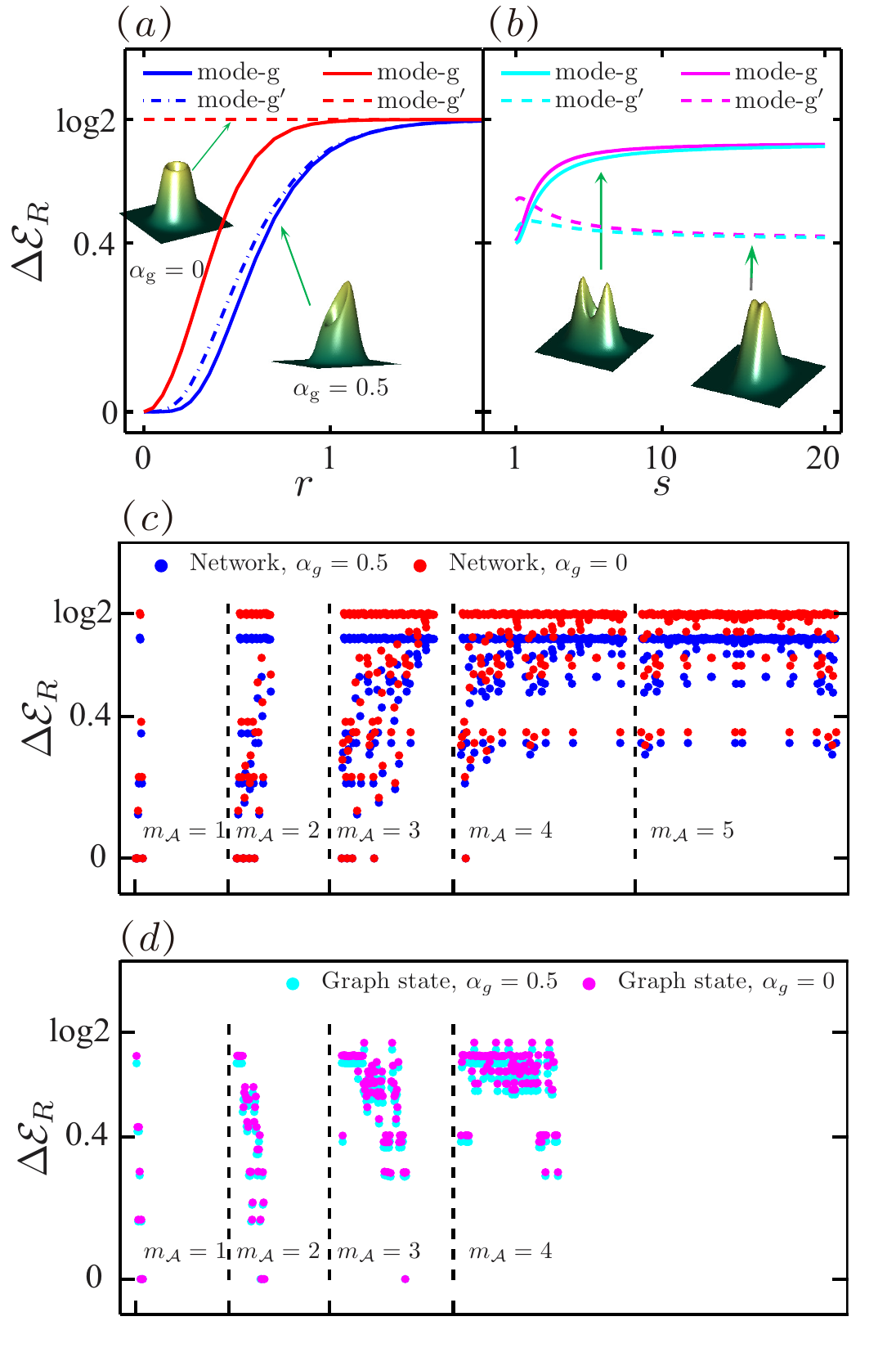}\\
\caption{\footnotesize (a) and (b) are the increase of entanglement ${\Delta\cal E}_{R}$ as the degree of squeezing changes,  corresponding to the network and graph states in Fig. \ref{fig2}, respectively. The solid  and dash curves  correspond to the photon-subtracted  mode (mode-g) and its neighboring mode (mode-$g^{\prime}$), respectively.  The blue (cyan)  and the red (pink) curves correspond to the small displacement  $\alpha_{g} = 0.5$ and $\alpha_{g} = 0$, respectively.  The single-mode Wigner functions are  shown for the  curves  pointed by the green thin arrow.  
 (c) and (d) Each scattered dot represents a different bi-partition with $m_{\cal A}$ modes in subsystem $\cal A$. 
 The blue and the red scattered dots in Panel (c) correspond to  the first example  in Fig. \ref{fig2}(a). The cyan and the pink scattered dots in Panel(d) correspond to the second example in Fig. \ref{fig2}(b).
} \label{fig3}
\end{figure}

In Fig.~\ref{fig2}, each circle represents a mode in a multimode quantum state, and the color code inside the circle indicate  the increase in entanglement between the mode (subsystem $\mathcal A$) and the rest system (subsystem $\mathcal B$).  When a photon is subtracted from the mode $g$ ( marked by the green arrow) in the pure multimode systems, non-Gaussian features  will spread from the mode $g$ to other modes,  whose entanglement  can also increase.  
 In Fig.~\ref{fig3}(a) and Fig.~\ref{fig3}(b), we respectively show the increase of entanglement (${\Delta\cal E}_{R}$) related to mode $g$ and mode $g^{\prime}$ (representing the neighboring mode of $g$ and marked by a green circle in Fig.~\ref{fig2}). As shown by the blue and pink curves,
 ${\Delta\cal E}_{R}$ increases as the degree of squeezing increases until it approaches a maximum. The maximum values are usually different for different initial Gaussian states, but they are all less than $\log2$. In Figs.~\ref{fig3}(c) and \ref{fig3}(d), each dots represents a different bi-partition for the examples of Figs.~\ref{fig2}(a) and \ref{fig2}(2), respectively, which we group depending on the number of modes in subsystem $\mathcal A$. The entanglement increase varies depending on the modes that are comprised in each subsystem, but it is never greater than $\log2$.

Finally, we aim to provide an intuition for the value $\log2$ of entanglement increase. To this goal, we restrict to a two-mode system prepared in a non-displaced pure state $\ket{\psi} = \hat U (\ket{0}\otimes \ket{0})$, where $\hat U$ is a Bogoliubov transformation. When we subtract a photon in one of the two modes, we can use (\ref{parallelmain}) to write the photon-subtracted state $\ket{\psi^-} = \hat U \hat b (\ket{0}\otimes \ket{0})/{\cal N}$ (with ${\cal N}$ a normalisation factor). We recall that $\hat{b} = \hat U^{\dag} \hat a_g \hat U =\vec k \cdot \vec {\hat{a}}^{\dag}+\vec l \cdot \vec{\hat{a}}$, which implies that 
\begin{equation}
 \ket{\psi^-} = \frac{1}{\cal N}\hat U \left( k_1\ket{1}\otimes \ket{0} + k_2 \ket{0}\otimes \ket{1}\right).
\end{equation}
In particular in the limit where the squeezing induced by $\hat U$ is small, we find that $\ket{\psi}$ resembles the vacuum state with negligible entanglement. On the other hand, $\ket{\psi^-}$ is close to a single photon that passed through a beam splitter. The maximal entanglement in this low squeezing limit is achieved for the Bell state $\ket{\psi^-} \approx (\ket{1}\otimes \ket{0} \pm \ket{0}\otimes \ket{1})/\sqrt{2}$, which has a R\'enyi-2 entanglement ${\cal E}_R(\ket{\phi}) = \log 2$. This corresponds exactly to the maximal entanglement increase and thus we can interpret our findings as a generalisation of this phenomenon to mixed states of arbitrary many modes with arbitrary amounts of squeezing. Because of this connection to Bell states, the maximal entanglement increase $\Delta{\cal E}_R = \log 2$ may also be referred to as one $e$bit \cite{Bennett}.\\

\section{Discussion}

 We strictly prove that single-photon subtraction can reduce the purity of any multimode Gaussian state by at most a factor $1/2$. This means that for any  bipartite entanglement of all Gaussian pure states, the amount of R\'enyi-2 entanglement that can be gained by single-photon subtraction is at most $\log2$. We then show that mixed photon-subtracted Gaussian states can be decomposed as a mixture of pure photon subtracted states and that the upper bound on entanglement increase also applies to mixed states. A fully equivalent treatment of single-photon addition yields the same result (see end of Appendix \ref{B1}). Finally, we evaluate the actual R\'enyi-2 entanglement of two multimode examples based on the Wigner function and show that the bound can be reached. 

Furthermore, we highlight that eq.~(\ref{eq:decompEnt1}) can more generally be used to evaluate state-dependent upper bounds for the increase in R\'enyi-2 entanglement due to photon subtracted states. It can thus be used to evaluate the robustness of the additional non-Gaussian entanglement to photon loss using the methods of \cite{WalschaersRa}. Furthermore, photon subtraction is also a method to remotely generate Wigner negativity through Einstein-Podolsky-Rosen Steering \cite{M. Walschaers PRL,M. Walschaers PRXQ}. As an outlook, we expect that the bound of eq.~(\ref{eq:decompEnt1}) can help understand whether there is a trade-off between the generated Wigner negativity and the increase in non-Gaussian entanglement. This highlights the importance of our results in understanding the resourcefulness of non-Gaussian operations and their potential use in quantum technologies. \\

{\em Acknowledgements---} K.Z. and J.J. acknowledge financial support through the Innovation Program of Shanghai Municipal Education Commission (Grant No. 2021-01-07-00-08-E00100); the National Natural Science Foundation of China (11874155, 91436211, 11374104, 12174110); Basic Research Project of Shanghai Science and Technology Commission (20JC1416100); Natural Science Foundation of Shanghai (17ZR1442900); Minhang Leading Talents (201971); Shanghai Sailing Program (21YF1410800); Shanghai Municipal Science and Technology Major Project (2019SHZDZX01); the 111 project (B12024).
N.T. and M.W. received funding from the European Union’s Horizon 2020 research and innovation programme under grant agreement No 899587. M.W. is also supported by the ANR JCJC project NoRdiC (ANR-21-CE47-0005).\\

 \bibliographystyle{plain}

\begin{thebibliography}{9}
\expandafter\ifx\csname natexlab\endcsname\relax\def\natexlab#1{#1}\fi
\expandafter\ifx\csname bibnamefont\endcsname\relax
  \def\bibnamefont#1{#1}\fi
\expandafter\ifx\csname bibfnamefont\endcsname\relax
  \def\bibfnamefont#1{#1}\fi
\expandafter\ifx\csname citenamefont\endcsname\relax
  \def\citenamefont#1{#1}\fi
\expandafter\ifx\csname url\endcsname\relax
  \def\url#1{\texttt{#1}}\fi
\expandafter\ifx\csname urlprefix\endcsname\relax\def\urlprefix{URL }\fi
\providecommand{\bibinfo}[2]{#2}
\providecommand{\eprint}[2][]{\url{#2}}
\bibitem{AEinstein}
\bibinfo{author}{\bibfnamefont{A.} \bibnamefont{Einstein}},
\bibinfo{author}{\bibfnamefont{B.} \bibnamefont{Podolsky}},
 and \bibinfo{author}{\bibfnamefont{N.} \bibnamefont{ Rosen}},
  \href{https://doi.org/10.1103/PhysRev.47.777}{
  \bibinfo{journal}{Phys. Rev. A} \textbf{\bibinfo{volume}{47}},
  \bibinfo{pages}{777} (\bibinfo{year}{1935}).} 
\bibitem{V. Giovannetti}
\bibinfo{author}{\bibfnamefont{V. Giovannetti}},
\bibinfo{author}{\bibfnamefont{S. Lloyd}},
and \bibinfo{author}{\bibfnamefont{L. Maccone}},\href{https://doi.org/10.1103/PhysRevLett.96.010401}{
  \bibinfo{journal}{Phys. Rev. Lett.} \textbf{\bibinfo{volume}{96}},
  \bibinfo{pages}{010401} (\bibinfo{year}{2006}).}
\bibitem{C. H. Bennett2}
\bibinfo{author}{\bibfnamefont{C. H. Bennett}},
\bibinfo{author}{\bibfnamefont{G. Brassard}},
\bibinfo{author}{\bibfnamefont{C. Cr\'{e}peau}},
\bibinfo{author}{\bibfnamefont{R. Jozsa}},
\bibinfo{author}{\bibfnamefont{A. Peres}},
and \bibinfo{author}{\bibfnamefont{W. K. Wootters}},
\href{https://doi.org/10.1103/PhysRevLett.70.1895}{  \bibinfo{journal}{Phys. Rev. Lett.} \textbf{\bibinfo{volume}{70}},
  \bibinfo{pages}{1895} (\bibinfo{year}{1993}).}
\bibitem{P. W. Shor}
\bibinfo{author}{\bibfnamefont{P. W. Shor}},\href{https://doi.org/10.1103/PhysRevA.52.R2493}{
  \bibinfo{journal}{Phys. Rev.  A} \textbf{\bibinfo{volume}{52}},
  \bibinfo{pages}{R2493} (\bibinfo{year}{1995}).}
\bibitem{S. Steinlechner}
\bibinfo{author}{\bibfnamefont{S. Steinlechner}},
\bibinfo{author}{\bibfnamefont{J. Bauchrowitz}},
\bibinfo{author}{\bibfnamefont{M. Meinders}},
\bibinfo{author}{\bibfnamefont{H. Müller-Ebhardt}},
\bibinfo{author}{\bibfnamefont{K. Danzmann}},
\bibinfo{author}{\bibfnamefont{R. Schnabel}},
\href{https://doi.org/10.1038/nphoton.2013.150}{
  \bibinfo{journal}{Nat. Photon.} \textbf{\bibinfo{volume}{7}},
  \bibinfo{pages}{626} (\bibinfo{year}{2013}).}
   \bibitem{B. Schumacher}
\bibinfo{author}{\bibfnamefont{B. Schumacher}},
and \bibinfo{author}{\bibfnamefont{M. A. Nielsen}},
 \href{https://doi.org/10.1103/PhysRevA.54.2629}{ \bibinfo{journal}{Phys. Rev. A} \textbf{\bibinfo{volume}{54}},
  \bibinfo{pages}{2629} (\bibinfo{year}{1996}).}
   \bibitem{S. Lloyd}
\bibinfo{author}{\bibfnamefont{S. Lloyd}},
\href{https://doi.org/10.1103/PhysRevA.55.1613}{
  \bibinfo{journal}{Phys. Rev. A} \textbf{\bibinfo{volume}{55}},
  \bibinfo{pages}{1613} (\bibinfo{year}{1997}).}


   \bibitem{R. Raussendorf}
\bibinfo{author}{\bibfnamefont{R. Raussendorf}},
and \bibinfo{author}{\bibfnamefont{H. J. Briegel}},
 \href{https://doi.org/10.1103/PhysRevLett.86.5188}{ \bibinfo{journal}{Phys. Rev. Lett.} \textbf{\bibinfo{volume}{86}},
  \bibinfo{pages}{5188} (\bibinfo{year}{2001}).}
   \bibitem{E. Knill}
\bibinfo{author}{\bibfnamefont{E. Knill}},
\bibinfo{author}{\bibfnamefont{R. Laflamme}},
and \bibinfo{author}{\bibfnamefont{G. J. Milburn}},
\href{https://doi.org/10.1038/35051009}{  \bibinfo{journal}{Nature} \textbf{\bibinfo{volume}{409}},
  \bibinfo{pages}{46} (\bibinfo{year}{2001}).}
\bibitem{S. L. Braunstein}
\bibinfo{author}{\bibfnamefont{S.} \bibfnamefont{L.} \bibnamefont{Braunstein}}
 and \bibinfo{author}{\bibfnamefont{P.} \bibfnamefont{van} \bibnamefont{Loock}},
\href{https://doi.org/10.1103/RevModPhys.77.513}{\bibinfo{journal}{Rev. Mod. Phys.} \textbf{\bibinfo{volume}{77}},
  \bibinfo{pages}{513} (\bibinfo{year}{2005}).}

  \bibitem{X. Su}
\bibinfo{author}{\bibfnamefont{X. Su}},
\bibinfo{author}{\bibfnamefont{Y. Zhao}},
\bibinfo{author}{\bibfnamefont{S. Hao}},
\bibinfo{author}{\bibfnamefont{X. Jia}},
\bibinfo{author}{\bibfnamefont{C. Xie}},
  and \bibinfo{author}{\bibfnamefont{K. Peng}},
\href{https://doi.org/10.1364/OL.37.005178}{\bibinfo{journal}{Opt. Lett.} \textbf{\bibinfo{volume}{37}},
  \bibinfo{pages}{5178} (\bibinfo{year}{2012}).}
\bibitem{S. Armstrong}
\bibinfo{author}{\bibfnamefont{S. Armstrong}},
\bibinfo{author}{\bibfnamefont{J.-F. Morizur}},
\bibinfo{author}{\bibfnamefont{J. Janousek}},
\bibinfo{author}{\bibfnamefont{B. Hage}},
\bibinfo{author}{\bibfnamefont{N. Treps}},
\bibinfo{author}{\bibfnamefont{P. K. Lam}},
and \bibinfo{author}{\bibfnamefont{H.-A. Bachor}},
\href{https://doi.org/10.1038/ncomms2033}{\bibinfo{journal}{Nat. Commun.} \textbf{\bibinfo{volume}{3}},
  \bibinfo{pages}{1206} (\bibinfo{year}{2012}).}
  \bibitem{M. Chen}
\bibinfo{author}{\bibfnamefont{M. Chen}},
\bibinfo{author}{\bibfnamefont{N. C. Menicucci}},
  and \bibinfo{author}{\bibfnamefont{O. Pfister}},
\href{https://doi.org/10.1103/PhysRevLett.112.120505}{\bibinfo{journal}{Phys. Rev. Lett.} \textbf{\bibinfo{volume}{112}},
  \bibinfo{pages}{120505} (\bibinfo{year}{2014}).}
  \bibitem{J. Roslund}
\bibinfo{author}{\bibfnamefont{J. Roslund}},
\bibinfo{author}{\bibfnamefont{R. M. de Ara\'{u}jo}},
\bibinfo{author}{\bibfnamefont{S. Jiang}},
\bibinfo{author}{\bibfnamefont{C. Fabre}},
and \bibinfo{author}{\bibfnamefont{N. Treps}},
\href{https://doi.org/10.1038/nphoton.2013.340}{\bibinfo{journal}{ Nat. Photon.} \textbf{\bibinfo{volume}{8}},
  \bibinfo{pages}{109} (\bibinfo{year}{2014}).}
  \bibitem{S. Gerke}
\bibinfo{author}{\bibfnamefont{S. Gerke}},
\bibinfo{author}{\bibfnamefont{J. Sperling}},
\bibinfo{author}{\bibfnamefont{W. Vogel}},
\bibinfo{author}{\bibfnamefont{Y. Cai}},
\bibinfo{author}{\bibfnamefont{J. Roslund}},
\bibinfo{author}{\bibfnamefont{N. Treps}},
and \bibinfo{author}{\bibfnamefont{C. Fabre}},
\href{https://doi.org/10.1103/PhysRevLett.114.050501}{\bibinfo{journal}{Phys. Rev. Lett.} \textbf{\bibinfo{volume}{114}},
  \bibinfo{pages}{050501} (\bibinfo{year}{2015}).}
  \bibitem{J.-i. Yoshikawa}
\bibinfo{author}{\bibfnamefont{J.-i. Yoshikawa}},
\bibinfo{author}{\bibfnamefont{S. Yokoyama}},
\bibinfo{author}{\bibfnamefont{T. Kaji}},
\bibinfo{author}{\bibfnamefont{C. Sornphiphatphong}},
\bibinfo{author}{\bibfnamefont{Y. Shiozawa}},
\bibinfo{author}{\bibfnamefont{K. Makino}},
and \bibinfo{author}{\bibfnamefont{A. Furusawa}},
\href{https://doi.org/10.1063/1.4962732}{\bibinfo{journal}{APL Photonics} \textbf{\bibinfo{volume}{1}},
  \bibinfo{pages}{060801} (\bibinfo{year}{2016}).}
\bibitem{K. Zhang}
\bibinfo{author}{\bibfnamefont{K. Zhang}},
\bibinfo{author}{\bibfnamefont{W. Wang}},
\bibinfo{author}{\bibfnamefont{S. Liu}},
\bibinfo{author}{\bibfnamefont{X. Pan}},
\bibinfo{author}{\bibfnamefont{J. Du}},
\bibinfo{author}{\bibfnamefont{Y. Lou}},
\bibinfo{author}{\bibfnamefont{S. Yu}},
\bibinfo{author}{\bibfnamefont{S. Lv}},
\bibinfo{author}{\bibfnamefont{N. Treps}},
\bibinfo{author}{\bibfnamefont{C. Fabre}},
and \bibinfo{author}{\bibfnamefont{J. Jing}},
\href{https://doi.org/10.1103/PhysRevLett.124.090501}{\bibinfo{journal}{Phys. Rev. Lett.} \textbf{\bibinfo{volume}{124}},
  \bibinfo{pages}{090501} (\bibinfo{year}{2020}).}
\bibitem{W. Wang}
\bibinfo{author}{\bibfnamefont{W. Wang}},
\bibinfo{author}{\bibfnamefont{K. Zhang}},
and \bibinfo{author}{\bibfnamefont{J. Jing}},
\href{https://doi.org/10.1103/PhysRevLett.125.140501}{\bibinfo{journal}{Phys. Rev. Lett.} \textbf{\bibinfo{volume}{125}},
  \bibinfo{pages}{140501} (\bibinfo{year}{2020}).}
\bibitem{S. D. Bartlett}
\bibinfo{author}{\bibfnamefont{S. D. Bartlett}},
\bibinfo{author}{\bibfnamefont{B. C. Sanders}},
\bibinfo{author}{\bibfnamefont{S. L. Braunstein}},
and \bibinfo{author}{\bibfnamefont{K. Nemoto}},
\href{https://doi.org/10.1103/PhysRevLett.88.097904}{\bibinfo{journal}{Phys. Rev. Lett.} \textbf{\bibinfo{volume}{88}},
  \bibinfo{pages}{097904} (\bibinfo{year}{2002}).}
\bibitem{A. Mari}
\bibinfo{author}{\bibfnamefont{A. Mari}}
and \bibinfo{author}{\bibfnamefont{J. Eisert}},
\href{https://doi.org/10.1103/PhysRevLett.109.230503}{\bibinfo{journal}{Phys. Rev. Lett.} \textbf{\bibinfo{volume}{109}},
  \bibinfo{pages}{230503} (\bibinfo{year}{2012}).}
\bibitem{S. Rahimi-Keshari}
\bibinfo{author}{\bibfnamefont{S. Rahimi-Keshari}},
\bibinfo{author}{\bibfnamefont{T. C. Ralph}},
and \bibinfo{author}{\bibfnamefont{C. M. Caves}},
\href{https://doi.org/10.1103/PhysRevX.6.021039}{\bibinfo{journal}{Phys. Rev. X} \textbf{\bibinfo{volume}{6}},
  \bibinfo{pages}{021039} (\bibinfo{year}{2016}).}
   \bibitem{A. Zavatta}
\bibinfo{author}{\bibfnamefont{A. Zavatta}},
\bibinfo{author}{\bibfnamefont{V. Parigi}},
and \bibinfo{author}{\bibfnamefont{M. Bellini}},
 \href{https://doi.org/10.1103/PhysRevA.75.052106}{ \bibinfo{journal}{Phys. Rev. A} \textbf{\bibinfo{volume}{75}},
  \bibinfo{pages}{052106} (\bibinfo{year}{2007}).}
   \bibitem{J. Wenger}
\bibinfo{author}{\bibfnamefont{J. Wenger}},
\bibinfo{author}{\bibfnamefont{R. Tualle-Brouri}},
and \bibinfo{author}{\bibfnamefont{P. Grangier}},
 \href{https://doi.org/10.1103/PhysRevLett.92.153601}{ \bibinfo{journal}{Phys. Rev. Lett.} \textbf{\bibinfo{volume}{92}},
  \bibinfo{pages}{153601} (\bibinfo{year}{2004}).}
   \bibitem{A. Ourjoumtsev}
\bibinfo{author}{\bibfnamefont{A. Ourjoumtsev}},
\bibinfo{author}{\bibfnamefont{A. Dantan}},
 \bibinfo{author}{\bibfnamefont{R. Tualle-Brouri}},
and \bibinfo{author}{\bibfnamefont{P. Grangier}},
  \href{https://doi.org/10.1103/PhysRevLett.98.030502}{\bibinfo{journal}{Phys. Rev. Lett.} \textbf{\bibinfo{volume}{98}},
  \bibinfo{pages}{030502} (\bibinfo{year}{2007}).}
   \bibitem{V. Parigi}
\bibinfo{author}{\bibfnamefont{V. Parigi}},
\bibinfo{author}{\bibfnamefont{A. Zavatta}},
 \bibinfo{author}{\bibfnamefont{M. Kim}},
and \bibinfo{author}{\bibfnamefont{M. Bellini}},
  \href{https://doi.org/10.1126/science.1146204}{\bibinfo{journal}{Science} \textbf{\bibinfo{volume}{317}},
  \bibinfo{pages}{1890} (\bibinfo{year}{2007}).}
   \bibitem{Y.-S. Ra}
\bibinfo{author}{\bibfnamefont{Y.-S. Ra}},
\bibinfo{author}{\bibfnamefont{A. Dufour}},
\bibinfo{author}{\bibfnamefont{M. Walschaers}},
\bibinfo{author}{\bibfnamefont{C. Jacquard}},
\bibinfo{author}{\bibfnamefont{T. Michel}},
\bibinfo{author}{\bibfnamefont{C. Fabre}},
and \bibinfo{author}{\bibfnamefont{N. Treps}},
  \href{https://doi.org/10.1038/s41567-019-0726-y}{\bibinfo{journal}{Nat. Phys.} \textbf{\bibinfo{volume}{16}},
  \bibinfo{pages}{144 - 147} (\bibinfo{year}{2020}).}
   \bibitem{Cochrane}
\bibinfo{author}{\bibfnamefont{P. T. Cochrane}},
\bibinfo{author}{\bibfnamefont{T. C. Ralph}},
and \bibinfo{author}{\bibfnamefont{G. J. Milburn}},
 \href{https://doi.org/10.1103/PhysRevA.65.062306}{ \bibinfo{journal}{Phys. Rev. A} \textbf{\bibinfo{volume}{65}},
  \bibinfo{pages}{062306} (\bibinfo{year}{2002}).}
   \bibitem{Olivares}
\bibinfo{author}{\bibfnamefont{S. Olivares}},
\bibinfo{author}{\bibfnamefont{M. G. A. Paris}},
and \bibinfo{author}{\bibfnamefont{R. Bonifacio}},
  \href{https://doi.org/10.1103/PhysRevA.67.032314}{\bibinfo{journal}{Phys. Rev. A} \textbf{\bibinfo{volume}{67}},
  \bibinfo{pages}{032314} (\bibinfo{year}{2003}).}
   \bibitem{Yang Yang}
\bibinfo{author}{\bibfnamefont{Y. Yang}},
and \bibinfo{author}{\bibfnamefont{F.-L. Li}},
  \href{https://doi.org/10.1103/PhysRevA.80.022315}{\bibinfo{journal}{Phys. Rev. A} \textbf{\bibinfo{volume}{80}},
  \bibinfo{pages}{022315} (\bibinfo{year}{2009}).}
   \bibitem{Carlos Navarrete-Benlloch}
\bibinfo{author}{\bibfnamefont{C. Navarrete-Benlloch}},
\bibinfo{author}{\bibfnamefont{R. Garc\'{i}a-Patr\'{o}n}},
\bibinfo{author}{\bibfnamefont{J. H. Shapiro}},
and \bibinfo{author}{\bibfnamefont{N. J. Cerf}},
 \href{https://doi.org/10.1103/PhysRevA.86.012328}{ \bibinfo{journal}{Phys. Rev. A} \textbf{\bibinfo{volume}{86}},
  \bibinfo{pages}{012328} (\bibinfo{year}{2012}).}
   \bibitem{M. Walschaers2}
\bibinfo{author}{\bibfnamefont{M.} \bibnamefont{Walschaers}},
\bibinfo{author}{\bibfnamefont{C.} \bibnamefont{Fabre}},
\bibinfo{author}{\bibfnamefont{V.} \bibnamefont{Parigi}},
and \bibinfo{author}{\bibfnamefont{N.} \bibnamefont{Treps}},
 \href{https://doi.org/10.1103/PhysRevLett.119.183601}{ \bibinfo{journal}{Phys. Rev. Lett.} \textbf{\bibinfo{volume}{119}},
  \bibinfo{pages}{183601} (\bibinfo{year}{2017}).}
   \bibitem{M. Walschaers1}
\bibinfo{author}{\bibfnamefont{M.} \bibnamefont{Walschaers}},
\bibinfo{author}{\bibfnamefont{C.} \bibnamefont{Fabre}},
\bibinfo{author}{\bibfnamefont{V.} \bibnamefont{Parigi}},
and \bibinfo{author}{\bibfnamefont{N.} \bibnamefont{Treps}},
  \href{https://doi.org/10.1103/PhysRevA.96.053835}{\bibinfo{journal}{Phys. Rev. A.} \textbf{\bibinfo{volume}{96}},
  \bibinfo{pages}{053835} (\bibinfo{year}{2017}).}
   \bibitem{M. Walschaers3}
\bibinfo{author}{\bibfnamefont{M.} \bibnamefont{Walschaers}},
\bibinfo{author}{\bibfnamefont{S. } \bibnamefont{Sarkar}},
\bibinfo{author}{\bibfnamefont{V.} \bibnamefont{Parigi}},
and \bibinfo{author}{\bibfnamefont{N.} \bibnamefont{Treps}},
 \href{https://doi.org/10.1103/PhysRevLett.121.220501}{ \bibinfo{journal}{Phys. Rev. Lett.} \textbf{\bibinfo{volume}{121}},
  \bibinfo{pages}{220501} (\bibinfo{year}{2018}).}
   \bibitem{NoGo1}
   \bibinfo{author}{ \bibnamefont{J. Eisert}},
\bibinfo{author}{ \bibnamefont{S. Scheel}}, and
\bibinfo{author}{\bibnamefont{M. B. Plenio}},
  \href{https://doi.org/10.1103/PhysRevLett.89.137903}{\bibinfo{journal}{Phys. Rev. Lett.} \textbf{\bibinfo{volume}{89}},
  \bibinfo{pages}{137903} (\bibinfo{year}{2002}).}
   \bibitem{NoGo2}
\bibinfo{author}{\bibnamefont{J. Fiur\'a\ifmmode \check{s}\else \v{s}\fi{}ek}},
 \href{https://doi.org/10.1103/PhysRevLett.89.137904}{ \bibinfo{journal}{Phys. Rev. Lett.} \textbf{\bibinfo{volume}{89}},
  \bibinfo{pages}{137904} (\bibinfo{year}{2002}).}
   \bibitem{NoGo3}
\bibinfo{author}{\bibnamefont{G. Giedke}},
and \bibinfo{author}{\bibnamefont{J. I. Cirac}}
  \href{https://doi.org/10.1103/PhysRevA.66.032316}{\bibinfo{journal}{Phys. Rev. A} \textbf{\bibinfo{volume}{66}},
  \bibinfo{pages}{032316} (\bibinfo{year}{2002}).}
   \bibitem{EntDist}
\bibinfo{author}{\bibnamefont{H. Takahashi}},
\bibinfo{author}{\bibnamefont{J. S. Neergaard-Nielsen}},
\bibinfo{author}{\bibnamefont{M. Takeuchi}},
\bibinfo{author}{\bibnamefont{M. Takeoka}},
\bibinfo{author}{\bibnamefont{K. Hayasaka}},
\bibinfo{author}{\bibnamefont{A. Furusawa}},
and \bibinfo{author}{\bibnamefont{M. Sasaki}}
 \href{https://doi.org/10.1038/nphoton.2010.1}{ \bibinfo{journal}{Nat. Photon.} \textbf{\bibinfo{volume}{4}},
  \bibinfo{pages}{178–181} (\bibinfo{year}{2010}).}
\bibitem{Vidal-Werner}
\bibinfo{author}{\bibfnamefont{G. Vidal}},
  and \bibinfo{author}{\bibfnamefont{R. F. Werner}},
  \href{https://doi.org/10.1103/PhysRevA.65.032314}{\bibinfo{journal}{Phys. Rev. A} \textbf{\bibinfo{volume}{65}},
  \bibinfo{pages}{032314} (\bibinfo{year}{2002}).}
\bibitem{R. Horodecki}
\bibinfo{author}{\bibfnamefont{R. Horodecki}},
\bibinfo{author}{\bibfnamefont{P. Horodecki}},
\bibinfo{author}{\bibfnamefont{M. Horodecki}},
  and \bibinfo{author}{\bibfnamefont{K. Horodecki}},
  \href{https://doi.org/10.1103/RevModPhys.81.865}{\bibinfo{journal}{Rev. Mod. Phys.} \textbf{\bibinfo{volume}{81}},
  \bibinfo{pages}{865} (\bibinfo{year}{2009}).}
   \bibitem{C. H. Bennett}
\bibinfo{author}{\bibfnamefont{C. H. Bennett}},
\bibinfo{author}{\bibfnamefont{D. P. DiVincenzo}},
\bibinfo{author}{\bibfnamefont{R. F.Werner}},
and \bibinfo{author}{\bibfnamefont{W. K.Wootters}},
\href{https://doi.org/10.1103/PhysRevA.54.3824}{  \bibinfo{journal}{Phys. Rev. A} \textbf{\bibinfo{volume}{54}},
  \bibinfo{pages}{3824} (\bibinfo{year}{1996}).}
   \bibitem{G. Toth}
   \bibinfo{author}{\bibfnamefont{G. T\'{o}th}},
\bibinfo{author}{\bibfnamefont{T. Moroder}},
and \bibinfo{author}{\bibfnamefont{O. G\"{u}hne}},
  \href{https://doi.org/10.1103/PhysRevLett.114.160501}{\bibinfo{journal}{Phys. Rev. Lett.} \textbf{\bibinfo{volume}{114}},
  \bibinfo{pages}{160501} (\bibinfo{year}{2015}).}
\bibitem{Kim_2010}
\bibinfo{author}{\bibfnamefont{J. S.} \bibnamefont{Kim}},
         and \bibinfo{author}{\bibfnamefont{B. C.} \bibnamefont{ Sanders}},
  \href{https://doi.org/10.1088/1751-8113/43/44/445305}{\bibinfo{journal}{J. Phys. A: Math. Theor.} \textbf{\bibinfo{volume}{43}},
  \bibinfo{pages}{445305} (\bibinfo{year}{2010}).}
   \bibitem{G. Adesso1}
\bibinfo{author}{\bibfnamefont{G. Adesso}},
and \bibinfo{author}{\bibfnamefont{A. Serafini}},
 \href{https://doi.org/10.1103/PhysRevLett.109.190502}{ \bibinfo{journal}{Phys. Rev. Lett.} \textbf{\bibinfo{volume}{109}},
  \bibinfo{pages}{190502} (\bibinfo{year}{2012}).}
   \bibitem{Lami1}
\bibinfo{author}{\bibfnamefont{L. Lami}},
\bibinfo{author}{\bibfnamefont{C. Hirche}},
\bibinfo{author}{\bibfnamefont{G. Adesso}},
and \bibinfo{author}{\bibfnamefont{A. Winter}},
 \href{https://doi.org/10.1103/PhysRevLett.117.220502}{ \bibinfo{journal}{Phys. Rev. Lett.} \textbf{\bibinfo{volume}{117}},
  \bibinfo{pages}{220502} (\bibinfo{year}{2016}).}
   \bibitem{Lami2}
\bibinfo{author}{\bibfnamefont{L. Lami}},
\bibinfo{author}{\bibfnamefont{L. Mi\v{s}ta, Jr.}},
and \bibinfo{author}{\bibfnamefont{G. Adesso}},
 \href{https://doi.org/10.48550/arXiv.2010.15729}{ \bibinfo{journal}{ArXiv} 
  \bibinfo{pages}{2010.15729} (\bibinfo{year}{2020}).}
     \bibitem{ModesStates}
\bibinfo{author}{\bibfnamefont{C. Fabre}}
and \bibinfo{author}{\bibfnamefont{N. Treps}},
 \href{https://doi.org/10.1103/RevModPhys.92.035005}{ \bibinfo{journal}{Rev. Mod. Phys.}
 \textbf{\bibinfo{volume}{92}}
  \bibinfo{pages}{035005} (\bibinfo{year}{2020}).}
     \bibitem{Tutorial}
\bibinfo{author}{\bibfnamefont{M. Walschaers}},
 \href{https://doi.org/10.1103/PRXQuantum.2.030204}{ \bibinfo{journal}{PRX Quantum}
 \textbf{\bibinfo{volume}{2}}
  \bibinfo{pages}{030204} (\bibinfo{year}{2021}).}
\bibitem{C. Weedbrook}
\bibinfo{author}{\bibfnamefont{C.} \bibnamefont{Weedbrook}},
\bibinfo{author}{\bibfnamefont{S.} \bibnamefont{Pirandola}},
\bibinfo{author}{\bibfnamefont{R.} \bibnamefont{ García-Patrón}},
 \bibinfo{author}{\bibfnamefont{N. J.} \bibnamefont{ Cerf}},
 \bibinfo{author}{\bibfnamefont{T. C.} \bibnamefont{ Ralph}},
\bibinfo{author}{\bibfnamefont{J. H. } \bibnamefont{Shapiro}},
         and \bibinfo{author}{\bibfnamefont{S.} \bibnamefont{ Ralph}},
  \href{https://doi.org/10.1103/RevModPhys.84.621}{\bibinfo{journal}{Rev. Mod. Phys} \textbf{\bibinfo{volume}{84}},
  \bibinfo{pages}{621} (\bibinfo{year}{2012}).}
   \bibitem{M. M.Wolf}
\bibinfo{author}{\bibfnamefont{M. M.Wolf}},
\bibinfo{author}{\bibfnamefont{G. Giedke}},
\bibinfo{author}{\bibfnamefont{O. Kr\"{u}ger}},
\bibinfo{author}{\bibfnamefont{R. F.Werner}},
and \bibinfo{author}{\bibfnamefont{J. I. Cirac}},
 \href{https://doi.org/10.1103/PhysRevA.69.052320}{ \bibinfo{journal}{Phys. Rev. A} \textbf{\bibinfo{volume}{69}},
  \bibinfo{pages}{052320} (\bibinfo{year}{2004}).}
   \bibitem{Bennett}
\bibinfo{author}{\bibfnamefont{C. H. Bennett}},
\bibinfo{author}{\bibfnamefont{H. J. Bernstein}},
\bibinfo{author}{\bibfnamefont{S. Popescu}},
and \bibinfo{author}{\bibfnamefont{B. Schumacher}},
 \href{https://doi.org/10.1103/PhysRevA.53.2046}{ \bibinfo{journal}{Phys. Rev. A} \textbf{\bibinfo{volume}{53}},
  \bibinfo{pages}{2046} (\bibinfo{year}{1996}).}
   \bibitem{WalschaersRa}
\bibinfo{author}{\bibfnamefont{M. Walschaers}},
\bibinfo{author}{\bibfnamefont{Y.-S. Ra}},
and \bibinfo{author}{\bibfnamefont{N. Treps}},
  \href{https://doi.org/10.1103/PhysRevA.100.023828}{\bibinfo{journal}{Phys. Rev. A} \textbf{\bibinfo{volume}{100}},
  \bibinfo{pages}{023828} (\bibinfo{year}{2019}).}
   \bibitem{M. Walschaers PRL}
\bibinfo{author}{\bibfnamefont{M. Walschaers}},
and \bibinfo{author}{\bibfnamefont{N. Treps}},
 \href{https://doi.org/10.1103/PhysRevLett.124.150501}{ \bibinfo{journal}{Phys. Rev. Lett.} \textbf{\bibinfo{volume}{124}},
  \bibinfo{pages}{150501} (\bibinfo{year}{2020}).}
   \bibitem{M. Walschaers PRXQ}
\bibinfo{author}{\bibfnamefont{M. Walschaers}},
\bibinfo{author}{\bibfnamefont{V. Parigi}},
\bibinfo{author}{\bibfnamefont{N. Treps}},
\href{https://doi.org/10.1103/PRXQuantum.1.020305}{  \bibinfo{journal}{PRX Quantum} \textbf{\bibinfo{volume}{1}},
  \bibinfo{pages}{020305} (\bibinfo{year}{2020}).}
  
\end{thebibliography}

\appendix
\onecolumngrid
\section{Decomposition of single-photon subtraction operation}\label{A1}
In CV quantum optics,  
for an arbitrary $m$-mode Gaussian  state $\hat{\rho}_{0}$, 
we can write a thermal decomposition as
\begin{eqnarray}\label{00}
\hat{\rho}_{0}=\mathcal{\hat{D}}\hat{U}\hat{\rho}_{M}\hat{U}^{\dag}\mathcal{\hat{D}}^{\dag},
\end{eqnarray}
where $\hat{U}$ is an arbitrary canonical unitary and $\mathcal{\hat{D}}=\prod\limits_{i=1}^{m}\hat{D}(\alpha_{i})$ is displacement operator, which displace the vacuum state to generate coherent states $|\alpha_{i}\rangle=\hat{D}(\alpha_{i})|0\rangle$.  Furthermore,  $\hat{\rho}_M=\bigotimes\limits_{i=1}^{m}\hat{\rho}_i$, where  single-
mode thermal states $\hat{\rho}_i=(0,V_i)$ can be fully characterised by a covariance matrix  $V_i=\mathrm{diag}(n_i,n_i)$, , where $n_i$  is the fraction of thermal noise compared to shot noise. Then the purity of $\hat{\rho}_{0}(\bar{\mathrm{x}},V)$  can  be expressed as $\mu_{0} =\prod\limits_{i=1}^{m}\mathrm{tr}(\hat{\rho}_i^{2})= \prod\limits_{i=1}^{m}\frac{1}{n_i}$.


When subtracting a photon to  the mode-$\mathrm{g}$, it corresponds to a  unitary Bogoliubov transformation
\begin{eqnarray}
\hat{U}^{\dag}\mathcal{\hat{D}}^{\dag}\hat{a}_{\mathrm{g}}\mathcal{\hat{D}}\hat{U}&=& \hat{b}+\alpha_{\mathrm{g}}^{\ast}, \quad\mathrm{with}\quad \hat{b}=\vec k \cdot \vec {\hat{a}}^{\dag}+\vec l \cdot \vec{\hat{a}},
\end{eqnarray}
where $\vec {\hat{a}}^{\dag}=(\hat{a}_{1}^{\dag},\cdots,\hat{a}_{m}^{\dag})^{\top}$  and $\vec{\hat{a}}=(\hat{a}_{1},\cdots,\hat{a}_{m})^{\top}$ . A set of $1\times m$ complex matrices $\vec k=(k_{1},\cdots,k_{m})$ and $\vec l=(l_{1},\cdots,l_{m})$ corresponds an canonical unitary $\hat{U}$.    Thus, we   convert the single-photon subtraction operation on the  mode-$\mathrm{g}$ into the addition and subtraction operation on each thermal mode. Thus the single-photon subtracted state can then be written as
\begin{eqnarray}\label{parallel}
\frac{\hat{a}_{\mathrm{g}}\hat{\rho}_{0}\hat{a}^{\dag}_{\mathrm{g}}}{\mathrm{tr}(\hat{a}_{\mathrm{g}}\hat{\rho}_{0}\hat{a}^{\dag}_{\mathrm{g}})}=\frac{\mathcal{\hat{D}}\hat{U}(\hat{b}+\alpha_{\mathrm{g}})\hat{\rho}_{M}(\hat{b}^{\dag}+\alpha^{\ast}_{\mathrm{g}})\hat{U}^{\dag}\mathcal{\hat{D}}^{\dag}}{\mathcal{N}},
\end{eqnarray}
where the normalization factor $\mathcal{N}=\mathrm{tr}(\hat{b}\hat{\rho}_{M}\hat{b}^{\dag})+|\alpha_{\mathrm{g}}|^2$.
 Due to $\mathcal{\hat{D}}\hat{U}$ will not change the purity, so the purity of an arbitrary single-photon subtracted state can be expressed as
\begin{eqnarray}\label{um}
\nonumber \mu_{}{\mathcal{N}^2}&=&\mathrm{tr}([\hat{b}\hat{\rho}_{M}\hat{b}^{\dag}+|\alpha_{\mathrm{g}}|^2\hat{\rho}_{M}+\alpha_{\mathrm{g}}\hat{\rho}_{M}\hat{b}^{\dag}+\alpha_{\mathrm{g}}^{\ast}\hat{b}\hat{\rho}_{M}]^2)\\
\nonumber &=&\mathrm{tr}((\hat{b}\hat{\rho}_{M}\hat{b}^{\dag}+|\alpha_{\mathrm{g}}|^2\hat{\rho}_{M})^2+(\alpha_{\mathrm{g}}\hat{\rho}_{M}\hat{b}^{\dag}+\alpha^{\ast}_{\mathrm{g}}\hat{b}\hat{\rho}_{M})^2)\\
\nonumber &=&\mathrm{tr}(\hat{b}\hat{\rho}_{M}\hat{b}^{\dag}\hat{b}\hat{\rho}_{M}\hat{b}^{\dag})+|\alpha_{\mathrm{g}}|^4\mathrm{tr}(\hat{\rho}_{M}^2)+{\alpha_{\mathrm{g}}}^2\mathrm{tr}(\hat{\rho}_{M}\hat{b}^{\dag}\hat{\rho}_{M}\hat{b}^{\dag})+{\alpha^{\ast}_{\mathrm{g}}}^2\mathrm{tr}(\hat{b}\hat{\rho}_{M}\hat{b}\hat{\rho}_{M})+\\
   &&|\alpha_{\mathrm{g}}|^2[\mathrm{tr}(\hat{\rho}_{M}\hat{b}\hat{\rho}_{M}\hat{b}^{\dag})+\mathrm{tr}(\hat{b}\hat{\rho}_{M}\hat{b}^{\dag}\hat{\rho}_{M})]+|\alpha_{\mathrm{g}}|^2[\mathrm{tr}(\hat{\rho}_{M}\hat{b}^{\dag}\hat{b}\hat{\rho}_{M})+\mathrm{tr}(\hat{b}\hat{\rho}_{M}\hat{\rho}_{M}\hat{b}^{\dag})].
\end{eqnarray}
where   we have used   that $\mathrm{tr}(\hat{b}\hat{\rho}_{M}\hat{b}^{\dag}\hat{\rho}_{M}\hat{b}^{\dag})=\mathrm{tr}(\hat{b}\hat{\rho}_{M}\hat{b}^{\dag}\hat{b}\hat{\rho}_{M})=
\mathrm{tr}(\hat{\rho}_{M}\hat{\rho}_{M}\hat{b}^{\dag})=\mathrm{tr}(\hat{\rho}_{M}\hat{b}\hat{\rho}_{M})=0$, since $\hat{\rho}_{i}$ is a diagonal matrix,  and $\hat{a}_{i}^{\dag}$ $(\hat{a}_{i})$  is off-diagonal matrices.
Similarly, for single-photon addition, the calculation is equivalent the same expression just by swapping the values of $\vec k$ and $\vec l$ and the values of $\alpha_{\mathrm{g}}^{\ast}$ with $\alpha_{\mathrm{g}}$.

For the same Gaussian state, different  form  Eq.(\ref{00}),  we can also take the following decomposition
\begin{eqnarray}
\hat{\rho}_{0}=\hat{U}\mathcal{\hat{D}}(\bar{\mathrm{x}})\hat{\rho}_{M}\mathcal{\hat{D}}(\bar{\mathrm{x}})^{\dag}\hat{U}^{\dag}.
\end{eqnarray}
In this case,  we just need to to replace $\alpha_{\mathrm{g}}^{\ast}$ with $\sum_{i=1}^{m}k_i\alpha^{\ast}+l_i\alpha$ in Eq.(\ref{um}), which  is still applicable, regardless of the decomposition form.  

\section{Generalized expression of relative purity}\label{B1}

\subsection{When $\alpha_{\mathrm{g}}=0$}
When $\alpha_{\mathrm{g}}=0$,  only the first term of (\ref{um}) is retained.

Let us take $m=1$, so that (\ref{um}) becomes
\begin{eqnarray}
\mu_{1}{N}_{1}^{2}=&&\mathrm{tr}(\hat{b}\hat{\rho}_{M}\hat{b}^{\dag}\hat{b}\hat{\rho}_{M}\hat{b}^{\dag})\\
\nonumber =&&|k_1|^4\mathrm{tr}(\hat{a}_{1}^{\dag}\hat{\rho}_{1}\hat{a}_{1}\hat{a}_{1}^{\dag}\hat{\rho}_{1}\hat{a}_{1})+|l_1|^4\mathrm{tr}(\hat{a}_{1}\hat{\rho}_{1}\hat{a}_{1}^{\dag}\hat{a}_{1}\hat{\rho}_{1}\hat{a}_{1}^{\dag})
+k_{1}^2l_{1}^{\ast2}\mathrm{tr}(\hat{a}_{1}^{\dag}\hat{\rho}_{1}\hat{a}_{1}^{\dag}\hat{a}_{1}^{\dag}\hat{\rho}_{1}\hat{a}_{1}^{\dag})+k_{1}^{\ast2}l_{1}^2\mathrm{tr}(\hat{a}_{1}\hat{\rho}_{1}\hat{a}_{1}\hat{a}_{1}\hat{\rho}_{1}\hat{a}_{1})\\
\nonumber &&+ |k_1|^2|l_{1}|^{2}[\mathrm{tr}(\hat{a}_{1}^{\dag}\hat{\rho}_{1}\hat{a}_{1}\hat{a}_{1}\hat{\rho}_{1}\hat{a}_{1}^{\dag})+\mathrm{tr}(\hat{a}_{1}\hat{\rho}_{1}\hat{a}_{1}^{\dag}\hat{a}_{1}^{\dag}\hat{\rho}_{1}\hat{a}_{1}) ]+|k_1|^2|l_{1}|^{2}[\mathrm{tr}(\hat{a}_{1}^{\dag}\hat{\rho}_{1}\hat{a}_{1}^{\dag}\hat{a}_{1}\hat{\rho}_{1}\hat{a}_{1})+\mathrm{tr}(\hat{a}_{1}\hat{\rho}_{1}\hat{a}_{1}\hat{a}_{1}^{\dag}\hat{\rho}_{1}\hat{a}_{1}^{\dag})]\\  \nonumber &&+|k_1|^2 k_{1}l_{1}^{\ast}[\mathrm{tr}(\hat{a}_{1}^{\dag}\hat{\rho}_{1}\hat{a}_{1}\hat{a}_{1}^{\dag}\hat{\rho}_{1}\hat{a}_{1}^{\dag})+\mathrm{tr}(\hat{a}_{1}^{\dag}\hat{\rho}_{1}\hat{a}_{1}^{\dag}\hat{a}_{1}^{\dag}\hat{\rho}_{1}\hat{a}_{1})]
+|k_1|^2 k_{1}^{\ast}l_{1}[\mathrm{tr}(\hat{a}_{1}^{\dag}\hat{\rho}_{1}\hat{a}_{1}\hat{a}_{1}\hat{\rho}_{1}\hat{a}_{1})+\mathrm{tr}(\hat{a}_{1}\hat{\rho}_{1}\hat{a}_{1}\hat{a}_{1}^{\dag}\hat{\rho}_{1}\hat{a}_{1})]\\
&&+|l_1|^2 k_{1}l_{1}^{\ast}[\mathrm{tr}(\hat{a}_{1}\hat{\rho}_{1}\hat{a}_{1}^{\dag}\hat{a}_{1}^{\dag}\hat{\rho}_{1}\hat{a}_{1}^{\dag})+\mathrm{tr}(\hat{a}_{1}^{\dag}\hat{\rho}_{1}\hat{a}_{1}^{\dag}\hat{a}_{1}\hat{\rho}_{1}\hat{a}_{1}^{\dag})]
+|l_1|^2 k_{1}^{\ast}l_{1}[\mathrm{tr}(\hat{a}_{1}\hat{\rho}_{1}\hat{a}_{1}^{\dag}\hat{a}_{1}\hat{\rho}_{1}\hat{a}_{1})+\mathrm{tr}(\hat{a}_{1}\hat{\rho}_{1}\hat{a}_{1}\hat{a}_{1}\hat{\rho}_{1}\hat{a}_{1}^{\dag})].\nonumber
\end{eqnarray}
Remove the zero traces and replace the subscript 1 with $i$, we get any single-mode purity
\begin{eqnarray}
\mu_{i}{N}_{i}^{2}=&&|k_i|^4\mathrm{tr}(\hat{a}_{i}^{\dag}\hat{\rho}_{i}\hat{a}_{i}\hat{a}_{i}^{\dag}\hat{\rho}_{i}\hat{a}_{i})+|l_i|^4\mathrm{tr}(\hat{a}_{i}\hat{\rho}_{i}\hat{a}_{i}^{\dag}\hat{a}_{i}\hat{\rho}_{i}\hat{a}_{i}^{\dag})
\\&&\nonumber+ 2|k_i|^2|l_{i}|^{2}\mathrm{tr}(\hat{a}_{i}^{\dag}\hat{\rho}_{i}\hat{a}_{i}\hat{a}_{i}\hat{\rho}_{i}\hat{a}_{i}^{\dag})+2|k_i|^2|l_{i}|^{2}\mathrm{tr}(\hat{a}_{i}^{\dag}\hat{\rho}_{i}\hat{a}_{i}^{\dag}\hat{a}_{i}\hat{\rho}_{1}\hat{a}_{i}). \end{eqnarray}
When we generalise this to a state with an arbitrary number of modes $m$, we find that the purity can be expressed as
\begin{eqnarray}\label{umm}
\nonumber \mu_{m}(\sum_{i=1}^{m}{N}_{i})^{2}=&&\sum_{i=1}^{m}\mu_{i}{N}_{i}^{2}\cdot n_i\cdot\prod_{i=1}^{m}\frac{1}{n_i}+4\cdot[|\sum_{i=1}^{m}[k_il_i\mathrm{tr}(\hat{a}_{i}\hat{\rho}_{i}\hat{a}_{i}^{\dag}\hat{\rho}_{i})\cdot n_{i}|^{2}-\sum_{i=1}^{m}|k_il_i\mathrm{tr}(\hat{a}_{i}\hat{\rho}_{i}\hat{a}_{i}^{\dag}\hat{\rho}_{i})\cdot n_{i}|^2]\cdot\prod_{i=1}^{m}\frac{1}{n_i}+\\
\nonumber &&[(\sum_{i=1}^{m}[|k_i|^2\mathrm{tr}(\hat{a}_{i}^{\dag}\hat{\rho}_{i}^{2}\hat{a}_{i})+|l_i|^2\mathrm{tr}(\hat{a}_{i}\hat{\rho}_{i}^{2}\hat{a}_{i}^{\dag})]\cdot n_{i})^{2}-\sum_{i=1}^{m}[|k_i|^2\mathrm{tr}(\hat{a}_{i}^{\dag}\hat{\rho}_{i}^{2}\hat{a}_{i})+|l_i|^2\mathrm{tr}(\hat{a}_{i}\hat{\rho}_{i}^{2}\hat{a}_{i}^{\dag})]^2\cdot n_{i}^{2}]\cdot\prod_{i=1}^{m}\frac{1}{n_i}+\\
&&[(\sum_{i=1}^{m}[(|k_i|^2+|l_i|^2)\mathrm{tr}(\hat{a}_{i}\hat{\rho}_{i}\hat{a}_{i}^{\dag}\hat{\rho}_{i})]\cdot n_{i})^{2}-\sum_{i=1}^{m}[(|k_i|^2+|l_i|^2)\mathrm{tr}(\hat{a}_{i}\hat{\rho}_{i}\hat{a}_{i}^{\dag}\hat{\rho}_{i})]^2\cdot n_{i}^{2}]\cdot\prod_{i=1}^{m}\frac{1}{n_i},
\end{eqnarray}
where $\mathcal{N}_{i} =|k_{i}|^2\mathrm{tr}(\hat{a}_{i}^{\dag}\rho_{i} \hat{a}_{i})+|l_{i}|^2\mathrm{tr}(\hat{a}_{i}\rho_{i}\hat{a}_{i}^{\dag})$.
Therefore, the traces in (\ref{umm}) are  related to the single-mode photon-added and -subtracted thermal states, which are the key to deriving the expression of final purity. Fortunately, 
 Winger function provides us with an easy way to calculate these traces.  For a single-mode  Gaussian thermal state $\hat{\rho}_i$, the Wigner function   is
\begin{eqnarray}
 w_{i}(q_i,p_i) &=&\frac{1}{2\pi n_i} \mathrm{exp}[-\frac{q_{i}^2+p_{i}^2}{2n_i}],
\end{eqnarray}
where $q_i$ and $p_i$ are amplitude and phase quadrature. Then the Wigner functions due to photon addition or subtraction are
\begin{eqnarray}\label{Wng}
 w_{(i,+1)}(q_i,p_i) &=&(\frac{n_i+1}{2n_i}(q_{i}^2+p_{i}^2)-\frac{1}{n_i}) w_{i}(q_i,p_i), \\
 w_{(i,-1)}(q_i,p_i)&=&(\frac{n_i-1}{2n_i}(q_{i}^2+p_{i}^2)+\frac{1}{n_i}) w_{i}(q_i,p_i).
\end{eqnarray}
Thus, we can  calculate the  traces by the above  Wigner functions:
\begin{eqnarray}
  \mathrm{tr}(\hat{a}\hat{\rho}_{i}\hat{a}^{\dag}) &=&\int \frac{q_i^2+p_i^2}{4} w_{i}\mathrm{d}q_i\mathrm{d}p_i-1=\frac{n_{i}-1}{2},\\
  \mathrm{tr}(\hat{a}^{\dag}\hat{\rho}_{i}\hat{a}) &=&\int \frac{q_i^2+p_i^2}{4} w_{i}\mathrm{d}q_i\mathrm{d}p_i-1+1=\frac{n_{i}+1}{2},
\end{eqnarray}
 which are the quantities related to the average-photon in the  thermal state.  Similarly, we can get that 
\begin{eqnarray}
  \mathrm{tr}(\hat{a}\hat{\rho}_{i}\hat{a}^{\dag}\hat{a}\hat{\rho}_{i}\hat{a}^{\dag}) &=&\frac{\mathrm{tr}(\hat{a}\hat{\rho}_{i}\hat{a}^{\dag}\hat{a}\hat{\rho}_{i}\hat{a}^{\dag})}{\mathrm{tr}(\hat{a}\hat{\rho}_{i}\hat{a}^{\dag})^2}\cdot\mathrm{tr}(\hat{a}\hat{\rho}_{i}\hat{a}^{\dag})^2 =\frac{(1+n_{i}^2)}{2n_{i}^3}(\frac{n_{i}-1}{2})^2,\\
\mathrm{tr}(\hat{a}^{\dag}\hat{\rho}_{i}\hat{a}\hat{a}^{\dag}\hat{\rho}_{i}\hat{a}) &=&\frac{(1+n_{i}^2)}{2n_{i}^3}(\frac{n_{i}+1}{2})^2,
\end{eqnarray}
 which are quantities related to the purity of the photon-added and -subtracted states.
\begin{eqnarray}
\mathrm{tr}(\hat{a}^{\dag}\hat{\rho}_{i}\hat{a}\hat{a}\hat{\rho}_{i}\hat{a}^{\dag})&=&\mathrm{tr}(\hat{a}\hat{\rho}_{i}\hat{a}^{\dag}\hat{a}^{\dag}\hat{\rho}_{i}\hat{a})
 =\mathrm{tr}(\hat{\rho}_{i}\hat{a}^{\dag}\hat{a}^{\dag}\hat{\rho}_{i}\hat{a}\hat{a})
 =\frac{(n_{i}^2-1)^2}{8n_{i}^3},
\end{eqnarray}
which can  be obtained by photon-added and -subtracted Wigner functions through the following integration 
\begin{eqnarray}
\mathrm{tr}(\hat{a}^{\dag}\hat{\rho}_{i}\hat{a}\hat{a}\hat{\rho}_{i}\hat{a}^{\dag})&=&(4\pi\int w_{(i,+1)}\cdot w_{(i,-1)}\mathrm{d}q_i\mathrm{d}p_i)\cdot(\int \frac{q_i^2+p_i^2}{4} w_{i}\mathrm{d}q_i\mathrm{d}p_i-1)\cdot(\int \frac{q_i^2+p_i^2}{4} w_{i}\mathrm{d}q_i\mathrm{d}p_i)\\
 &=&\nonumber\frac{(n_{i}^2-1)^2}{8n_{i}^3}.
\end{eqnarray}
With the help of  Wigner function of  single-photon added or subtracted  and Gaussian thermal state Wigner function, we can get
\begin{eqnarray}
\nonumber \mathrm{tr}(\hat{\rho}_{i}\hat{a}^{\dag}\hat{\rho}_{i}\hat{a}) &=&\mathrm{tr}(\hat{\rho}_{i}\hat{a}\hat{\rho}_{i}\hat{a}^{\dag}) = (4\pi\int w_{i}\cdot w_{(i,-1)}\mathrm{d}q_i\mathrm{d}p_i)\cdot(\int \frac{q_i^2+p_i^2}{4} w_{i}\mathrm{d}q_i\mathrm{d}p_i-1)=\frac{n_{i}+1}{2n_{i}^2}(\frac{n_{i}-1}{2}).
\end{eqnarray}
Due to the noncommutative nature of creation and annihilation operators, we have
\begin{eqnarray}
  \mathrm{tr}(\hat{a}\hat{\rho}_{i}\hat{a}\hat{a}^{\dag}\hat{\rho}_{i}\hat{a}^{\dag}) =\mathrm{tr}(\hat{a}\hat{\rho}_{i}\hat{a}^{\dag}\hat{a}\hat{\rho}_{i}\hat{a}^{\dag})+\mathrm{tr}(\hat{\rho}_{i}^2\hat{a}^{\dag}\hat{a})
\nonumber =\mathrm{tr}(\hat{a}^{\dag}\hat{\rho}_{i}\hat{a}\hat{a}^{\dag}\hat{\rho}_{i}\hat{a})-\mathrm{tr}(\hat{\rho}_{i}^2\hat{a}^{\dag}\hat{a})-\mathrm{tr}(\hat{\rho}_{i}^2),
\end{eqnarray}
then we can deduce the following traces
\begin{eqnarray}
\mathrm{tr}(\hat{\rho}_{i}^2\hat{a}^{\dag}\hat{a}) &=& \frac{(n_{i}-1)^2}{4n_{i}^2},\quad \mathrm{tr}(\hat{\rho}_{i}^2\hat{a}\hat{a}^{\dag})=\frac{(n_{i}+1)^2}{4n_{i}^2}
\quad \mathrm{and}  \quad \mathrm{tr}(\hat{a}\hat{\rho}_{i}\hat{a}\hat{a}^{\dag}\hat{\rho}_{i}\hat{a}^{\dag}) =\frac{(n_{i}^2-1)^2}{8n_{i}^3}.
\end{eqnarray}

By bringing the results of these traces into (\ref{umm}), we get
\begin{eqnarray}
 \nonumber {\mu_{m}^{-}} =&&[\frac{1}{2}(\sum_{i=1}^{m}{N}_{i})^2+\frac{1}{2}(\sum_{i=1}^{m}\frac{\tilde{{N}}_{i}}{n_i})^2+|\sum_{i=1}^{m}k_{i}l_{i}\frac{n_i^2-1}{2n_{i}}|^2]{\mu_{0}} /{(\sum_{i=1}^{m}{N}_{i})^2}.
\end{eqnarray}
where ${N}_{i} =|k_{i}|^2\frac{n_i+1}{2}+|l_{i}|^2\frac{n_i-1}{2}$ and ${\tilde{N}}_{i} =|k_{i}|^2\frac{n_i+1}{2}-|l_{i}|^2\frac{n_i-1}{2}$. Hence, we can immediately know the relative purity  is
\begin{eqnarray}\label{RP}
 \nonumber \frac{\mu_{m}^{-}}{\mu_{0}}
 =&&\frac{1}{2}+[\frac{1}{2}(\sum_{i=1}^{m}\frac{\tilde{{N}}_{i}}{n_i})^2+|\sum_{i=1}^{m}k_{i}l_{i}\frac{n_i^2-1}{2n_{i}}|^2]/{(\sum_{i=1}^{m}{N}_{i})^2}\geqslant1/2.
\end{eqnarray}

\subsection{When $\alpha_{\mathrm{g}}\neq0$}

According to similar approach above, we can derive the following traces
\begin{eqnarray}
|\alpha_{\mathrm{g}}|^4\mathrm{tr}(\hat{\rho}_{M}^2)&=& |\alpha_{\mathrm{g}}|^4\mu_{0},\\
  {\alpha_{\mathrm{g}}}^2\mathrm{tr}(\hat{\rho}_{M}\hat{b}^{\dag}\hat{\rho}_{M}\hat{b}^{\dag}) &=& 2{\alpha_{\mathrm{g}}}^2(\sum_{i=1}^{m}k_{i}^{\ast}l_{i}^{\ast}\mathrm{tr}(\hat{a}_{i}\hat{\rho}_{i}\hat{a}_{i}^{\dag}\hat{\rho}_{i})n_i)\mu_{0}={\alpha_{\mathrm{g}}}^2(\sum_{i=1}^{m}\frac{k_{i}^{\ast}l_{i}^{\ast}(n_{i}^{2}-1)}{2n_{i}})\mu_{0},\\
  {\alpha^{\ast}_{\mathrm{g}}}^2\mathrm{tr}(\hat{b}\hat{\rho}_{M}\hat{b}\hat{\rho}_{M})&=&  2{\alpha^{\ast}_{\mathrm{g}}}^2(\sum_{i=1}^{m}k_{i}l_{i}\mathrm{tr}(\hat{a}_{i}\hat{\rho}_{i}\hat{a}_{i}^{\dag}\hat{\rho}_{i})n_i)\mu_{0}
  ={\alpha^{\ast}_{\mathrm{g}}}^2(\sum_{i=1}^{m}\frac{k_{i}l_{i}(n_{i}^{2}-1)}{2n_{i}})\mu_{0}
\end{eqnarray}
and
\begin{eqnarray}
\nonumber  &&|\alpha_{\mathrm{g}}|^2[\mathrm{tr}(\hat{\rho}_{M}\hat{b}\hat{\rho}_{M}\hat{b}^{\dag})+\mathrm{tr}(\hat{b}\hat{\rho}_{M}\hat{b}^{\dag}\hat{\rho}_{M})+\mathrm{tr}(\hat{\rho}_{M}\hat{b}^{\dag}\hat{b}\hat{\rho}_{M})+\mathrm{tr}(\hat{b}\hat{\rho}_{M}\hat{\rho}_{M}\hat{b}^{\dag})]\\
 \nonumber &&=2|\alpha_{\mathrm{g}}|^2[\sum_{i}^{m}(|k_{i}|^2+|l_{i}|^2)\mathrm{tr}(\hat{a}_{i}\hat{\rho}_{i}\hat{a}_{i}^{\dag}\hat{\rho}_{i})n_i+\sum_{i}^{m}(|k_{i}|^2\mathrm{tr}(\hat{\rho}_{i}^{2}\hat{a}_{i}\hat{a}_{i}^{\dag})n_i+|l_{i}|^2\mathrm{tr}(\hat{\rho}_{i}^{2}\hat{a}_{i}^{\dag}\hat{a}_{i})n_i)                        ]\mu_{0}\\
  &&=2|\alpha_{\mathrm{g}}|^2\sum_{i}^{m}N_i\mu_{0}.
\end{eqnarray}
Thus the relative purity is
\begin{eqnarray}\label{RP1}
 \nonumber \frac{\mu_{m}^{-}}{\mu_{0}}
 \nonumber =&&[\frac{1}{2}(\sum_{i=1}^{m}{N}_{i})^2+\frac{1}{2}(\sum_{i=1}^{m}\frac{\tilde{{N}}_{i}}{n_i})^2+2|\alpha_{\mathrm{g}}|^2\sum_{i=1}^{m}N_{i}+|\sum_{i=1}^{m}k_{i}l_{i}\frac{n_i^2-1}{2n_{i}}|^2+|\alpha_{\mathrm{g}}|^4+\\
 \nonumber &&{\alpha_{\mathrm{g}}}^2(\sum_{i=1}^{m}\frac{k_{i}^{\ast}l_{i}^{\ast}(n_{i}^{2}-1)}{2n_{i}})+{\alpha^{\ast}_{\mathrm{g}}}^2(\sum_{i=1}^{m}\frac{k_{i}l_{i}(n_{i}^{2}-1)}{2n_{i}})]/{(\sum_{i=1}^{m}{N}_{i}+|\alpha_{\mathrm{g}}|^2)^2}\\
 \nonumber   =&&\frac{1}{2}+[\frac{1}{2}(\sum_{i=1}^{m}\frac{\tilde{{N}}_{i}}{n_i})^2+
 \frac{1}{2}|\alpha_{\mathrm{g}}|^4+|\sum_{i=1}^{m}k_{i}l_{i}\frac{n_i^2-1}{2n_{i}}|^2\\
 &&|\alpha_{\mathrm{g}}|^2\sum_{i=1}^{m}N_{i}+{\alpha_{\mathrm{g}}}^2(\sum_{i=1}^{m}\frac{k_{i}^{\ast}l_{i}^{\ast}(n_{i}^{2}-1)}{2n_{i}})+{\alpha^{\ast}_{\mathrm{g}}}^2(\sum_{i=1}^{m}\frac{k_{i}l_{i}(n_{i}^{2}-1)}{2n_{i}})]/{(\sum_{i=1}^{m}{N}_{i}+|\alpha_{\mathrm{g}}|^2)^2}.
\end{eqnarray}
Since $(\alpha_{\mathrm{g}}+\alpha_{\mathrm{g}}^{\ast})^2\geqslant0$ and $(\alpha_{\mathrm{g}}-\alpha_{\mathrm{g}}^{\ast})^2\leqslant0$, thus we have  $-2|\alpha_{\mathrm{g}}|\leqslant \alpha_{\mathrm{g}}^2+{\alpha_{\mathrm{g}}^{\ast}}^2\leqslant2|\alpha_{\mathrm{g}}|^2$. Then
\begin{eqnarray}
B^2(\sum_{i=1}^{m}\frac{k_{i}^{\ast}l_{i}^{\ast}(n_{i}^{2}-1)}{2n_{i}})+{B^{\ast}}^2(\sum_{i=1}^{m}\frac{k_{i}l_{i}(n_{i}^{2}-1)}{2n_{i}})\geqslant-2|B|^2\sqrt{\sum_{i=1}^{m}\frac{k_{i}^{\ast}l_{i}^{\ast}(n_{i}^{2}-1)}{2n_{i}})}\cdot\sqrt{\sum_{i=1}^{m}\frac{k_{i}l_{i}(n_{i}^{2}-1)}{2n_{i}}},
\end{eqnarray}
where $B=\alpha_{\mathrm{g}}^{\ast}$. Furthermore, The left side of the equation has
\begin{eqnarray}
\nonumber  |\sum_{i=1}^{m}\frac{k_{i}l_{i}(n_{i}^{2}-1)}{2n_{i}}|^2 &=&\sum_{i=1}^{m}\frac{|k_{i}|^2|l_{i}|^2(n_{i}^{2}-1)^2}{(2n_{i})^2}+\sum_{i=1}^{m}\sum_{j=1}^{m}(k_{i}l_{i}{k_{j}}^{\ast}{l_{j}}^{\ast}+{k_{i}}^{\ast}{l_{i}}^{\ast}k_{j}l_{j})\frac{(n_{i}^{2}-1)}{2n_{i}}\frac{(n_{j}^{2}-1)}{2n_{j}}\\
\nonumber &\leqslant&\sum_{i=1}^{m}\frac{|k_{i}|^2|l_{i}|^2(n_{i}^{2}-1)^2}{(2n_{i})^2}+\sum_{i=1}^{m}\sum_{j=1}^{m}(2\sqrt{|k_{i}|^2|l_{i}|^2|k_{j}|^2|l_{j}|^2     })\frac{(n_{i}^{2}-1)}{2n_{i}}\frac{(n_{j}^{2}-1)}{2n_{j}}\\
&=&(\sum_{i=1}^{m} \sqrt{|k_{i}|^2|l_{i}|^2}\frac{(n_{i}^{2}-1)}{2n_{i}} )^2.
\end{eqnarray}
So that
\begin{eqnarray}\label{RP2}
B^2(\sum_{i=1}^{m}\frac{k_{i}^{\ast}l_{i}^{\ast}(n_{i}^{2}-1)}{2n_{i}})+{B^{\ast}}^2(\sum_{i=1}^{m}\frac{k_{i}l_{i}(n_{i}^{2}-1)}{2n_{i}})\geqslant-|B|^2\sum_{i=1}^{m} \sqrt{|k_{i}|^2|l_{i}|^2}\frac{(n_{i}^{2}-1)}{n_{i}},
\end{eqnarray}
and we add $|B|^2\sum_{i=1}^{m}N_{i}$ to both  sides of the equation 
\begin{eqnarray}
|B|^2&&\sum_{i=1}^{m}N_{i}+B^2(\sum_{i=1}^{m}\frac{k_{i}^{\ast}l_{i}^{\ast}(n_{i}^{2}-1)}{2n_{i}})+{B^{\ast}}^2(\sum_{i=1}^{m}\frac{k_{i}l_{i}(n_{i}^{2}-1)}{2n_{i}})\\&\geqslant&|B|^2(\sum_{i=1}^{m}N_{i}-\sum_{i=1}^{m} \sqrt{|k_{i}|^2|l_{i}|^2}\frac{(n_{i}^{2}-1)}{n_{i}}).\nonumber
\end{eqnarray}
Due to $\frac{n_i-1}{2}\geq0$, thus
\begin{eqnarray}
 \nonumber(\sum_{i=1}^{m}N_{i}-\sum_{i=1}^{m} &&\sqrt{|k_{i}|^2|l_{i}|^2}\frac{(n_{i}^{2}-1)}{n_{i}})\nonumber\\
 &=&\sum_{i=1}^{m}[(\sqrt{|k_{i}|^2\frac{n_i+1}{2}}-\sqrt{|l_{i}|^2\frac{n_i-1}{2}})^2+\sqrt{|k_{i}|^2|l_{i}|^2(n_i^2-1)}-\sqrt{|k_{i}|^2|l_{i}|^2}\frac{(n_{i}^{2}-1)}{n_{i}})]\nonumber\\
 \nonumber &=&\sum_{i=1}^{m}[(\sqrt{|k_{i}|^2\frac{n_i+1}{2}}-\sqrt{|l_{i}|^2\frac{n_i-1}{2}})^2+\sqrt{{|k_{i}|^2|l_{i}|^2(n_i^2-1)}}(\frac{\sqrt{n_i^{2}}-\sqrt{n_i^{2}-1}}{n_i})]\\
 &\geqslant& 0,
\end{eqnarray}
so that
\begin{eqnarray}
|B|^2\sum_{i=1}^{m}N_{i}+B^2(\sum_{i=1}^{m}\frac{k_{i}^{\ast}l_{i}^{\ast}(n_{i}^{2}-1)}{2n_{i}})+{B^{\ast}}^2(\sum_{i=1}^{m}\frac{k_{i}l_{i}(n_{i}^{2}-1)}{2n_{i}})\geqslant0.
  \end{eqnarray}
Therefore, the relative purity is always $1/2+$ a positive number, so that
\begin{eqnarray}
\frac{\mu_{m}^{-}}{\mu_{0}}\geqslant 1/2.
\end{eqnarray}
 It is a proof that for any Gaussian state, single-photon subtraction (or addition) can reduce the purity with at most  a factor 1/2.\\
 
 The case of photon addition yields the same expressions with $\alpha_g$ and $\alpha_g^{\ast}$ interchanged. In practice, the case of photon addition will also lead to different values for $k_i$ and $l_i$. However, this does not change any of the proofs and as a consequence, we also find $\mu_{m}^{+}/\mu_0 \geqslant 1/2$ for photon addition.




\end{document}